%% file: phmc_paper_1d.tex
\documentclass[12pt]{article}
\usepackage{epsfig}     

\newcommand{\nn}{\nonumber}

\newcommand\be{\begin{equation}}
\newcommand\ee{\end{equation}}
\newcommand\bea{\begin{eqnarray}}
\newcommand\eea{\end{eqnarray}}

\newcommand{\phd}{\phi^{\dagger}}

\def\ma[#1,#2,#3,#4]  {{\left( \matrix{ #1  & #2 \cr
                                        #3  & #4 \cr } \right)}}

\begin{document}

\title{
{\vspace{-0cm} \normalsize
\hfill \parbox{40mm}{CERN 98-237}\\
\hfill \parbox{40mm}{MPI-PhT/98-51}}\\[25mm]
The PHMC algorithm for \\simulations of dynamical fermions:\\
I -- description and properties}
\author{
Roberto  Frezzotti$^1$  
and Karl Jansen$^{2,}$\footnote{Heisenberg foundation fellow}
\\
{\footnotesize
$^1$ Max-Planck-Institut f\"ur Physik, F\"ohringer Ring 6, D-80805
     M\"unchen, Germany
}
\\
{\footnotesize 
$^2$ CERN, 1211 Gen\`eve 23, Switzerland  
}
}
\date{\today}
\maketitle

\begin{abstract}

We give a detailed description of the so-called
Polynomial Hybrid Monte Carlo (PHMC) algorithm.
The effects of the correction factor,
which is introduced to render the algorithm exact,
are discussed, stressing their relevance for the
statistical fluctuations and (almost) zero mode
contributions to physical observables.
We also investigate rounding-error effects and propose
several ways to reduce memory requirements. 

\end{abstract}

\input introduction_v2


\input section2_v4

\input section3_v6

\input section4_v2

\input section5_v2

\input conclusion

\input appendix_b

\input phmc.refs
\end{document}

%% file: introduction_v2
\section{Introduction}                     

Although lattice QCD \cite{Wilson} has nowadays reached a 
relatively mature age, precise quantitative results --at least in
the full theory-- are still rare. One of the main reasons is 
certainly that
numerical simulations of lattice QCD, including the
effects of dynamical quarks, are still very demanding and 
computer time consuming (for reviews of dynamical fermion
algorithms see \cite{forcrand3,karlreview}). 
Efforts to improve on this situation
are therefore highly desirable. 

In this paper we extend the discussion of the so-called
Polynomial Hybrid Monte Carlo (PHMC) algorithm, which we introduced
in \cite{frezzi} as an attempt to improve the performance
of simulation algorithms for dynamical fermions. 

The main idea of the PHMC 
algorithm relies on dividing  
the eigenvalue spectrum of the
Wilson-Dirac operator $M$ on the lattice into different disjoint parts. 
These different parts of the eigenvalue spectrum are then treated
by either incorporating them in the update step of a simulation algorithm 
or by taking them into account in a reweighing procedure. 
This general idea is realized in practise by designing suitable polynomials 
that approximate the inverse of $M^\dagger M$, which is needed in the
actual simulation, with a different accuracy   
for different parts of the eigenvalue spectrum of $M^\dagger M$. 
In the present paper we  choose a polynomial approximation to the inverse of
$M^\dagger M$ which is equivalent to basically neglecting
the contribution of the low-lying modes and taking very precisely
into account all the other modes in the update step. This choice, which follows the
original suggestion in \cite{taka}, 
is motivated by the experience with the multiboson technique 
\cite{lue94,bunk,peardon,beat,galli}:
neglecting in the update a small number of
low-lying modes of $M^\dagger M$ still yields results very close
to the ones obtained using the exact Hybrid Monte Carlo (HMC) algorithm.
We want to emphasize, however, that our choice 
is only a special case and the general method allows for a 
greater flexibility including ideas like the one proposed in \cite{duncan}. 

One may argue that the reweighing step can be replaced by a reject/accept
step in order to render the algorithm exact. We think that this is not the best choice
for the following reason:  
it is expected 
that almost zero modes
of the Wilson-Dirac operator appear when working in large physical volumes or
at large values of the lattice spacing.
In such a situation a reject/accept step leads to a
dilemma: either the acceptance probability becomes so small that such
events are always rejected. Or, if they are accepted, the zero modes
give exceptional values to quark propagators, distorting a statistical sample
substantially. However, in full QCD, gauge configurations carrying zero modes may
give a finite contribution to several fermion observables, which should
be taken into account --at least in principle-- in order to get correct
statistical averages. 
In fact, we consider this scenario as a potential danger 
for the Hybrid Monte Carlo (HMC) algorithm \cite{hmc} 
which is commonly used. 

As we will demonstrate below, with a suitable reweighing procedure, this
problem can be overcome elegantly. Namely, in our way of
correcting for the polynomial approximation, 
the reweighing factor becomes
proportional to the almost zero mode and hence cancels any singularity
appearing in 
quark propagators used to construct physical observables.
This mechanism reflects in a sense the role of the
determinant when the full QCD partition function is considered. 
Of course, reweighing techniques are widespread in
applications for numerical simulations. However, we would like to point out 
that our implementation of the reweighing factor
makes its computation very straightforward and reliable in all cases
and does not give too large an overhead in a simulation.  

In a previous publication \cite{frezzi} we introduced the PHMC algorithm 
and gave some first, promising results in practical applications for
Wilson fermions. 
However, it is by now well known 
that when using Wilson fermions for simulations of lattice QCD, 
one has to face large lattice cutoff effects 
in physical observables. For example, the axial Ward identity can
be substantially distorted in this case \cite{letter}.
However the effects of a non-vanishing lattice spacing can
be systematically reduced by applying Symanzik's improvement
programme \cite{symanzik}: this turns out to be easier in practice if only
on-shell quantities are to be ${\rm O}(a)$ improved \cite{on-shell}.  

In fact, implementing the improvement programme non-perturbatively
for both the action and all the local operators relevant for on-shell observables,
one can reach a complete cancellation of the cutoff effects that appear
linear in the lattice spacing \cite{paper1,paper3}. Since with such an 
improved theory we expect to be able to work at a much larger lattice spacing,
a substantial gain in the cost of numerical simulations can be obtained. 
Indeed, the non-perturbative on-shell ${\rm O}(a)$ improved action has by now already been
computed also for dynamical fermions \cite{jansom}. 
Any new simulation algorithm should hence have the ability to
be applicable to improved fermions. 
We therefore extend here our tests of the PHMC algorithm to the
case of ${\rm O}(a)$ improved actions. 
In the present paper we are going to discuss a number of 
important technical aspects of the PHMC algorithm.
Numerical results for the performance of the algorithm in practise
are deferred to a separate publication \cite{phmc_perf}.

%% file: section2_v4
\section{The PHMC algorithm}

We introduce the PHMC algorithm and discuss
several aspects concerning our practical implementation of the algorithm.               
In particular, we derive 
the computational cost of the algorithm in terms of matrix times vector
operations. 

\subsection{Introducing the PHMC algorithm}

We consider
Euclidean QCD with $n_f=2$ degenerate flavours regularized
on a hypercubic space-time lattice with lattice spacing $a$ and
size $L^3\times T$. With the lattice spacing set to unity from now on, the
points on the lattice have integer coordinates $(t,x_1,x_2,x_3)$ which
are in the range $0\le t \le T;0\le x_i < L$.
A gauge field $U_{\mu}(x)\in SU(3)$ is assigned to the link
pointing from the site $x$ to the site $(x+\mu)$, where
$\mu=0,1,2,3$ designates the 4 forward directions in space-time.
Throughout the paper we will adopt Schr\"odinger functional
boundary conditions as detailed in \cite{su3paper,sint,paper3}.
The partition function for lattice QCD with $n_f=2$ degenerate flavours
of quarks is given by 
\be \label{partitionfunction}
{\cal Z} = \int {\cal D}U
e^{- S_g[U]} \mbox{det}(Q^2[U]) \; ,
\ee
where $S_g$ is the standard Wilson-plaquette action for the pure gauge
sector with a coupling strength $\beta=6/g_0^2$ and $g_0$ the bare
gauge coupling. 
The Hermitean matrix $Q$, defining the fermion action, is given by    
\begin{eqnarray}
\label{qmatrix}
Q(U)_{xy} \!\!\!&=& \!\!\!\frac{c_0}{c_M}\gamma_5 [
(1+\sum_{\mu\nu}
[{i \over 2}c_{\rm sw}\kappa\sigma_{\mu\nu}{\cal F}_{\mu\nu}(x)])\delta_{x,y}
 \nonumber \\
&-&\kappa\sum_{\mu} \{
   (1-\gamma_{\mu})U_{\mu}(x)\delta_{x+\mu,y} +
(1+\gamma_{\mu})U^{\dagger}_{\mu}(x-\mu)\delta_{x-\mu,y}\}]  \;\;,
\end{eqnarray}
where $\kappa$ is the hopping parameter, related to the
bare quark mass $m_0$ by $\kappa=1/(8+2m_0)$ 
and $c_{\rm sw}$ is the ${\rm O}(a)$ improvement coefficient \cite{clover}.                                         
The constant 
$c_M$ serves to optimize 
simulation algorithms and 
$c_0=(1+8\kappa)^{-1}$.    
For all practical simulations we have imposed an even/odd preconditioning 
and hence used the preconditioned matrix $\hat{Q}$, whose precise definition 
can be found in e.g. \cite{janliu,rootorder}. 

It is the aim of the numerical simulations 
to compute expectation values of gauge invariant operators ${\cal O}$
\be \label{expectationvalue}
\langle {\cal O} \rangle = {\cal Z}^{-1}
                               \left[ \int {\cal D}U
e^{- S_g[U]} \mbox{det}(Q^2[U]) {\cal O}[U] \right]\; ,
\ee
using Monte Carlo methods. 
Note that in eq.(\ref{expectationvalue}) the square of the determinant
appears in order to have a positive definite measure suitable
for the numerical algorithms employed below. 

In the PHMC algorithm a polynomial $P_{n,\epsilon}(Q^2)$, 
approximating $(Q^2)^{-1}$ for all eigenvalues $\lambda$ of $Q^2$ with
$\lambda \in [\epsilon,1]$, is introduced such that
${\rm det}(P_{n,\epsilon}^{-1}(Q^2)) \approx {\rm det}(Q^2)$. 
Using the trivial identity ${\rm det}(Q^2)=
{\rm det}(Q^2P_{n,\epsilon}(Q^2))/{\rm det}(P_{n,\epsilon}(Q^2))$ and 
representing the determinants with the help of auxiliary bosonic fields
$\phi$ and $\eta$,
carrying  colour and spin indices, one may exactly rewrite the 
partition function eq.(\ref{partitionfunction}) as  
\bea \label{phmcpartition}
{\cal Z} &
 = & \int {\cal D}U{\cal D}\phd{\cal D}\phi
{\cal D}\eta^\dagger{\cal D}\eta
  \; W \; e^{- (S_g + S_P + S_\eta) } \nn \\
  S_P & = & S_P[U,\phi]=\phd P_{n,\epsilon}(Q^2[U]) \phi \nn \\
  S_\eta & = & \eta^\dagger\eta \; .
\eea
In eq.(\ref{phmcpartition}) we have introduced the ``correction factor'' 
$W=W[\eta,U]$:
\be \label{W_def}
W= \exp\left\{\eta^\dagger (1-[Q^2\cdot P_{n,\epsilon}(Q^2)]^{-1}) \eta \right\}
 \; .
\ee
Denoting averages evaluated with the effective action $S_g + S_P + S_\eta$
as $\langle \dots \rangle_P$, the exact averages denoted as 
$\langle \dots \rangle$ are obtained by reweighing with $W$ 
\be
\langle {\cal O} \rangle = \langle W \rangle_P^{-1}
\langle {\cal O}W \rangle_P  \label{true_ave} \; .
\ee   

As mentioned in the introduction, the advantage of rewriting the partition function 
in the form of eq.(\ref{phmcpartition}) is that by a suitable choice of
the polynomial $P_{n,\epsilon}(Q^2)$ the eigenvalue spectrum of $Q^2$ can be
smoothly separated into a part to be included 
in the update procedure by
simulating the effective action $S_g + S_P$ 
and a rest, taken into account in the correction factor. 

We remark that, in analogy to the case of the multiboson technique \cite{oneflavour},
the PHMC algorithm is also suited to allow for performing simulations
with an odd number of flavours. 
Of course, the above procedure leading to eq.(\ref{phmcpartition}) may be
generalized and several polynomials may be introduced in such a way
that each of them gives a good approximation in different parts of the
eigenvalue spectrum. We demand in this case that 
the product
of all these polynomials approximates the inverse of $Q^2$. 
The realization we are using in this paper amounts
to cutting out the very low-lying end of the eigenvalue spectrum from
the update step. 


In principle, there is a great flexibility in choosing the polynomial
to approximate $Q^{-2}$.            
In this work we follow ref.\cite{bunk} and
choose a Chebyshev approximation method to construct $P_{n,\epsilon}(Q^2)$. 
Since the polynomial we are going to use is detailed already 
in \cite{bunk,rootorder} we will give here just its final form
written in the 
product representation, 
\be \label{pol_prod_q}
P_{n,\epsilon}(Q^2) = p_{n,\epsilon}(Q) = \prod_{k=1}^{2n} [\sqrt{c_k}(Q-r_k)] \; ,
\ee
where the complex numbers $r_k$ are given by                  
\bea \label{roots_rk}
r_k & = & \sqrt{z_k} = \mu_k + i\nu_k \;, \; \nu_k > 0\; ,\;\;\; k=1,\dots ,n \nonumber \\ 
r_k & = & r_{2n+1-k}^{*} \; , \;\;\; k=n+1, \dots, 2n\; \nonumber \\
z_k & = & \frac{1}{2}(1+\epsilon) - \frac{1}{2}(1+\epsilon)\cos(\frac{2\pi k}{n+1})
    - i\sqrt{\epsilon}\sin(\frac{2\pi k}{n+1})\; .
\eea
The overall normalization constant, $\prod_{k=1}^n c_k$, can be computed
analytically. If the $c_k$'s are taken all identical, they turn out to be
of ${\rm O}(1)$.

The polynomial $P_{n,\epsilon}(Q^2)$ approximates the inverse of $Q^2$
with a relative fit error which is bounded from above by
\begin{equation} \label{accuracy}
\delta \equiv 2
\left(\frac{1-\sqrt{\epsilon}}{1+\sqrt{\epsilon}}\right)^{n+1}
\; ,
\end{equation}
for all the eigenmodes of $Q^2$ with eigenvalues $\lambda$ in the interval
$\lambda \in [\epsilon,1]$. The operator $Q^2$ is normalised (through
the choice of $c_M$) in such a way that its largest eigenvalue is always
smaller than $1$. 
For eigenvalues $\lambda < \epsilon$ the relative fit
error quickly increases as $\lambda$ decreases. 


\subsection{Implementation and cost of the PHMC algorithm}

The approximation of ${\rm det}(Q^2)$ through the 
inverse determinant of the polynomial $P_{n,\epsilon}(Q^2)$
was first suggested in \cite{lue94}. There it led to a completely local
bosonic action involving $n$ copies of bosonic fields. 
Since the bosonic action in that case was local, algorithms like
heatbath and over-relaxation could be used. One unfortunate property
of this approach was the observation that the autocorrelation time
of the algorithm grows with the number of bosonic fields appearing
in the action \cite{beat,galli}. 

The approximate fermion action $S_P$ eq.(\ref{phmcpartition}) 
\cite{taka,frezzi} 
on the other hand still represents a
non-local bosonic action.            
In this approach one therefore has to rely on small step size algorithms. 
However, 
the advantage is that now only one dynamical bosonic field is needed 
and hence
the dangerous increase of the autocorrelation time with the 
number of bosonic field copies mentioned above
is avoided. 

In more detail, we
have chosen to use a suitably adapted 
$\Phi$-version \cite{Phi_alg} of the HMC algorithm for
the update of the gauge fields. The usual arguments, including the reversibility 
of the molecular dynamics evolution, leading to the proof of detailed balance,
still holds for the case of the PHMC algorithm.                                               
The implementation of this update method for the case of 
${\rm O}(a)$ improved fermions and even/odd preconditioning 
can be done in complete analogy
to ref.~\cite{janliu}. We therefore only want to point out some peculiarities
which are not discussed in \cite{janliu}. 

In the following discussion we will be somewhat sketchy and focus our
attention on the modifications of the standard $\Phi$-version of the HMC algorithm
that are needed for implementing the PHMC algorithm. 
In particular, we note again that it is to be understood that for the actual simulation
the preconditioned matrix $\hat{Q}$ was always used. Another  remark is that 
the roots $r_k$, $k=1,\dots ,2n$ were suitably reordered with respect 
to their definition, eq.(\ref{roots_rk}), while preserving the
relation $r_{2n+1-k} = r_k^{*}$. Such a reordering is necessary to
keep rounding errors on a tolerable level, as thoroughly discussed 
in \cite{rootorder}. Details of the different ordering
schemes we have used in our implementation of the PHMC algorithm 
and rounding errors associated with them are discussed
in Section 4.

In the PHMC algorithm
the variation of the pseudofermion action, 
$S_P$ eq.(\ref{phmcpartition}), 
with respect to a given gauge link  
is somewhat more complicated than in the standard HMC algorithm. 
In terms of the variation of the operator $Q$, denoted by $\delta Q$,
it assumes the form
\be \label{PHMC_force}
\delta S_P
=  \sum_{j=1}^n 
                 \left[ \delta Q \; \chi_{j-1} \otimes \chi_{2n-j}^{\dagger}
               + \delta Q \; \chi_{2n-j} \otimes \chi_{j-1}^{\dagger} \right] \; ,
\ee
where the auxiliary pseudofermion fields $\chi_j$, for $j=1,\dots ,2n-1$ are
defined as
\be  \label{auxphi_def}
\chi_j \equiv [\sqrt{c_j}(Q-r_j)] \cdot [\sqrt{c_{j-1}}(Q-r_{j-1})]\cdot  
              \dots \cdot [\sqrt{c_1}(Q-r_1)] \phi 
\ee
and $\phi$ denotes the pseudofermion field of eq.(\ref{phmcpartition}).
In eq.(\ref{PHMC_force}) the products $\chi\otimes\chi^\dagger$ denote direct
products in colour space and a trace over spin indices is understood.

In order to speed up the simulation and minimize memory
requirements we proceed for the computation of $\delta S_P$ as follows:
We first precalculate the $n$ vectors 
$\chi_k$, for all $k=1,\dots, n$ and store them. 
We then start the evaluation of the different contributions to
$\delta S_P$ by computing $\chi_{n-1}\otimes\chi^\dagger_{n}$ 
and its Hermitean conjugate, which for brevity
will not be mentioned explicitly in the following.
The next contribution to $\delta S_P$ would involve
$\chi_{n-2}\otimes\chi^\dagger_{n+1}$. The vector 
$\chi^\dagger_{n+1}$ is obtained by computing $(Q-r_{n+1})\chi_{n}$.
The resulting vector can now be stored in $\chi_{n-1}$ 
since this vector is no longer
used. Iterating this procedure results in a memory 
requirement of $n+1$ pseudofermion vectors.
This may be considered as a drawback of the PHMC algorithm as it requires
a substantial amount of memory if the degree of the polynomial
becomes large. However, as it will be discussed below, there are
several ways to overcome possible bottlenecks if not enough memory is 
available.

It is clear that the evaluation of all terms necessary to evaluate 
$\delta S_P$ amounts to
$(2n-1)$ $Q\phi$ operations (the extra work to incorporate the
roots $r_k$ in the operator $Q-r_k$ is completely negligible). 
In addition, since there are $n$ terms to be summed (and traced) 
to evaluate eq.(\ref{PHMC_force}) and since each of them requires a
computational work roughly equivalent (at least in our implementation
on the APE computers) to one $Q\phi$ operation,
the complete cost of the computation of $\delta S_P$ 
will become about $3n$ $Q\phi$ operations.  

Although the polynomial
approximation to $(Q^2)^{-1}$ is rather precise even if a few eigenvalues
of $Q^2$ occur that are slightly larger than one, the numerical construction
of $\delta S_P$ itself turns out to be unstable when eigenvalues very
close to 1 are met in the updating procedure. At least in our implementation
of $\delta S_P$, based on eq.(\ref{PHMC_force}), numerical overflows
occurred when updating gauge configurations carrying modes of $Q^2$ with
eigenvalues very close to (even if smaller than) 1. 
In practise, $Q$ should therefore be normalized, through $c_M$ in eq.(\ref{qmatrix}), 
such that the average highest 
eigenvalue of $Q^2$ is sufficiently smaller than one, 
say $\langle \lambda_{\rm max}\rangle \approx 0.9$. 
Since the value of the highest eigenvalue
of $Q^2$ shows very small fluctuations, such an appropriate normalization can safely be
done at the beginning of a simulation.

The pseudofermion field $\phi$ in eq.(\ref{phmcpartition}) 
is to be generated according to the
distribution $\exp\left\{-S_P[U,\phi]\right\}$. 
Generating this distribution via a heatbath
step involves the computation of the inverse square root 
of $P_{n,\epsilon}(Q^2[U])$. 
This can be achieved by 
computing $\phi$ through 
\begin{equation}
\phi = A_{n,\epsilon}^\dagger(Q) [Q^2 P_{n,\epsilon}(Q^2)]^{-1} Q^2 R_G \nonumber 
\end{equation}
where $R_G$ is a random Gaussian vector and $A_{n,\epsilon}$ is given by
\bea \label{A}
A_{n,\epsilon}(Q) = \prod_{k=1}^{n} \sqrt{c_k} (Q-r_k) \; .
\eea
The vector $X=[Q^2 P_{n,\epsilon}(Q^2)]^{-1} Q^2 R_G$ 
is computed with a Conjugate Gradient (CG) method, solving the
equation $Q^2 P_{n,\epsilon}(Q^2) \; X = Q^2 R_G$. 
We demanded that in generating $\phi$ with a CG inverter
the relation 
\begin{equation} \label{check_phi}
| S_P - R_G^\dagger R_G | \approx {\rm O}(10^{-7})
\end{equation} 
holds. We noticed that this can be
achieved by choosing a moderately large stopping criterion for the CG solver,
namely $\epsilon_{\rm stop} = 10^{-12}$,
where $\epsilon_{\rm stop}$ is defined by the norm of the residual vector 
\be \label{residuum}
\Phi_{\rm res} = Q^2 R_G - Q^2 P_{n,\epsilon}(Q^2) \; X
\ee
divided by the norm of the solution vector $X$ (which is numerically close
to the norm of $Q^2 P_{n,\epsilon}(Q^2) \; X$ in all practical cases): 
\begin{equation} \label{eps_stop}
\epsilon_{\rm stop} = \|\Phi_{\rm res}\|^2 / \|X\|^2 .
\end{equation}

A last remark concerns the second pseudofermion field $\eta$. It is generated trivially 
by Gaussian random vectors. Through it the correction factor 
$W=W[\eta,U]$ (eq.(\ref{W_def})) can be computed 
via the solution of the equation 
$[Q^2 P_{n,\epsilon}(Q^2)]X = \eta$, 
which involves an additional inversion of $Q^2 P_{n,\epsilon}(Q^2)$.
The correction factor, or $w=\log(W)$ is then obtained by 
\be \label{W_eval}
 w[\eta,U] = \eta^\dagger (1-[Q^2 P_{n,\epsilon}(Q^2)]^{-1}) \eta\; . 
\ee
Since the expression $Q^2 P_{n,\epsilon}(Q^2)$ is almost the unit matrix, 
there is the possibility of dangerous rounding errors when computing 
the vector $(1-[Q^2 P_{n,\epsilon}(Q^2)]^{-1})\eta$ in 
eq.(\ref{W_eval}), especially on machines with only 32-bit precision
\footnote{We remark that, of course, all internal products and global
sums were performed in software Kahan or double precision arithmetic.}. 
However, eq.(\ref{W_eval}) may be rewritten as
\be \label{W_eval_R}
w[\eta,U] \equiv \eta^\dagger (R_{n,\epsilon}(Q^2)[Q^2 P_{n,\epsilon}(Q^2)]^{-1}) \eta \; .
\ee
Following refs.\cite{lue94,bunk,rootorder} the polynomial 
$R_{n,\epsilon}(Q^2) = Q^2 P_{n,\epsilon}(Q^2) - 1$ is
directly given by Chebyshev polynomials of degree $n+1$. One may hence use
numerically stable recursion relations 
to compute $R_{n,\epsilon}(Q^2)$.
Although the use of eq.~(\ref{W_eval_R}), instead of eq.~(\ref{W_eval}),  
leads to a somewhat larger cost for evaluating the correction factor,
our experience is that it is advisable to use eq.(\ref{W_eval_R}) when only 32-bit precision is
employed.
Analogously to the case of generating the pseudofermion field $\phi$,
eq.(\ref{phmcpartition}), we optimized the value of the stopping criterion
also for the CG inversion needed in eq.(\ref{W_eval_R}). 

It might be observed\footnote{We are grateful to Ulli Wolff for this
interesting remark.} that eq.(\ref{W_eval}) can be generalized to
\be \label{W_eval_gen}
 w[\eta,U] = \eta^\dagger (1-[b_{n,\epsilon}Q^2 P_{n,\epsilon}(Q^2)]^{-1}) \eta\; , 
\ee
where $b_{n,\epsilon}$ is some real positive constant. Its value might be
optimized, depending on the values of $n$ and $\epsilon$, in order
to reduce the stochastic noise associated with reweighing through the
correction factor. However, we did not exploit this additional freedom
and took always $b_{n,\epsilon}=1$, which enabled us to use
the expression of $w[\eta,U]$ in eq.(\ref{W_eval_R}).

In principle, the ratio of the number of $\eta$-field ``updates'' to
the number of gauge field updates is arbitrary. In fact, it turns out
(see Section 3)
that it is advantageous to choose this ratio
to be larger than one. In this way, the additional noise induced in 
the reweighted observables, eq.(\ref{true_ave}), by 
the correction factor can be partly suppressed. The above-mentioned ratio
will be denoted in the following by $N_{\rm corr}$, since it gives the number
of computations of the correction factor per gauge field configuration.  

From the discussion above it is easy to express the cost of the
PHMC algorithm in terms of matrix times vector, $Q\phi$, operations. 
The cost for the PHMC algorithm can be split into three parts,
\begin{equation} \label{PHMC_cost}
C_{Q\phi}(PHMC) = C_{\rm bhb} + C_{\rm update} + C_{\rm corr}\; ,
\end{equation}
where $C_{\rm bhb}$ is the cost for the heatbath of the bosonic fields,
$C_{\rm update}$ the cost for the computation of $\delta S_P$ and $C_{\rm corr}$ the cost
to evaluate the correction factor.
In units of $Q\phi$ operations we find
\begin{eqnarray} \label{PHMC_cost_comp}
C_{\rm bhb} & = & (2n+2)\cdot  N_{\rm CG}^{\rm bhb} + n\nonumber \\
C_{\rm update}  & = & 3n\cdot N_{\rm step} \nonumber \\
C_{\rm corr} & = & (2n+2)\cdot  N_{\rm CG}^{\rm corr} \cdot  N_{\rm corr}\; .
\end{eqnarray}
The factor $N_{\rm corr}$ denotes as above the number of evaluations
of the correction factor $W$ per full gauge field update (or molecular
dynamics trajectory). The symbols $N_{\rm CG}^{\rm bhb}$ and $N_{\rm CG}^{\rm corr}$
denote the average numbers of CG iterations in the heatbath of
the bosonic fields and the computation of $W$, respectively.
The factor $3n$ in $C_{\rm update}$ comes from adding the cost for the
construction of the auxiliary fields $\chi_k$ and the cost of the
other algebraic operations needed for a single update of the gauge
field and its conjugate momenta.                                   
$N_{\rm step}$ is the number of steps used in a trajectory, i.e.
how often $\delta S_P$ has to be evaluated within
a trajectory.
We explicitly verified that our formulae for $C_{\rm update}$, $C_{\rm bhb}$
and $C_{\rm corr}$ agree with the costs in real time observed for our
implementation of the PHMC algorithm on the APE computer.

The scaling behaviour of the computational cost $C_{\rm update}$, eq.(\ref{PHMC_cost_comp}),
as a function of the lattice size, $L^3 \times T$, or the condition number of $Q^2$ is 
expected to be fully analogous to the one observed in the HMC algorithm, with one
important difference. Due to the form of the variation of the pseudofermion action,
$S_P$ eq.(\ref{phmcpartition}), in the molecular dynamics evolution for the PHMC
algorithm the role of the lowest eigenvalue of $Q^2$ is taken over by the infrared
cut--off parameter of the polynomial approximation, $\epsilon$, as already discussed 
in \cite{frezzi}. 
Since in practise $\epsilon \approx 2\langle \lambda_{\rm min}\rangle$, we expect 
therefore an improvement on the cost of a simulation.

%% file: section3_v6
\section{Effects of the correction factor}

In this section we want to discuss the effects of the
correction factor we introduced for the exactness of
the algorithm. The first main point concerns the statistical
fluctuations induced by reweighing observables with
the correction factor: this aspect determines
to a large extent the tuning of the PHMC algorithm.
The second point is of qualitative nature and concerns
the occurrence of gauge configurations with exceptional
eigenvalues of $Q^2$ and how the reweighing procedure
can deal with them.

\subsection{Statistical errors and reweighing}

As discussed above, an important ingredient of the PHMC algorithm 
is the correction factor. The computational cost of the algorithm
will depend in a crucial way on the behaviour of the correction factor
in a real simulation. The reason is that through the correction
factor, which is computed stochastically, a certain 
noise is introduced which may affect the errors on the observables and
will contribute therefore to the cost of a simulation. 

In the PHMC algorithm 
the update of the gauge field $U$ is
alternated with $N_{\rm corr}$ ``updates'' of the pseudofermion field $\eta$, yielding
$N_{\rm corr}$ evaluations of $W[\eta,U]$ on each gauge configuration. 
Performing a simple arithmetic average of them yields a single estimate of the
correction factor per each gauge configuration. As a consequence, on a sample of $N$
gauge configurations labelled by the integer $j$, the averages $\langle \dots \rangle_P$
introduced in section 2.1 can be represented as trivial arithmetic 
averages over the sample:
\be \label{sample_ave}
\langle {\cal O} W \rangle_P = N^{-1} \sum_{j=1}^N {\cal O}_j W_j \;
\ee
where ${\cal O}_j$ is any gauge invariant observable and $W_j$ the above alluded
estimate of the correction factor on the gauge configuration $U_j$.

For any finite number of configurations, the statistical error on
$\langle {\cal O} \rangle$ is expected to depend on the choice of
$n$ and $\epsilon$, i.e on the chosen polynomial approximation
to $(Q^2)^{-1}$, and, for a given polynomial approximation, also
on the value of $N_{\rm corr}$.
 
By rewriting the expression for the reweighted average of ${\cal O}$,
eq.(\ref{true_ave}), in the form:
\be \label{true_ave_bis}
\langle {\cal O} \rangle = \langle {\cal O} \rangle_P +  \langle W \rangle_P^{-1} \cdot
( \langle {\cal O}W \rangle_P - \langle {\cal O} \rangle_P \langle W \rangle_P ) \; ,
\ee
it becomes clear that both the statistical fluctuations of ${\cal O}$ and those
of the connected part of ${\cal O} W$ contribute to the statistical error on the
reweighted average, eq.(\ref{true_ave_bis}). The latter contribution will depend 
on the statistical correlation between the observable ${\cal O}$ and the reweighing
factor $W$. 

As the polynomial approximation to $(Q^2)^{-1}$ is made more 
precise, the contribution to the error on ${\cal O}$ coming from the
statistical correlation between  ${\cal O}$ and $W$ becomes smaller,
just because $W$ gets closer to $1$.                    
Still, in the limit $W\approx 1$, there remains some noise, 
because 
$W$ is computed not exactly but only stochastically. 
We are then left with a pure Gaussian noise factor.                              
 
Choosing a poor polynomial approximation
to $(Q^2)^{-1}$ for the update step in the PHMC algorithm
yields a reweighing factor $W$
which will strongly fluctuate. (Think, e.g., of $W$ as being the full determinant.)
Moreover, $W$ may have in general a non-negligible correlation with the 
observable ${\cal O}$, such that, even in the limit $N_{\rm corr} \to \infty$,
a large contribution to the error on $\langle {\cal O} \rangle$ is expected
to arise. 
The discussion suggests that it is the relative statistical error of the correction 
factor itself that controls the additional fluctuations induced by reweighing and hence
the statistical error for a given observable $\langle {\cal O} \rangle$. 
 
As a consequence, we can expect the PHMC algorithm to be found efficient
only in situations where the variance of $W$ is very small. This amounts to choosing
the value of $\epsilon$ to be of the same order as the average lowest eigenvalue of
$(Q^2)$ and the value of $n$ to be large enough for the polynomial approximation
to $(Q^2)^{-1}$ to be reasonably precise. As we will see below, situations of this kind
can be realized, in practice, by setting $\epsilon\approx 2\langle \lambda_{\rm min}\rangle$ 
and $n$ such that the fit accuracy $\delta \approx 0.01$, see eq.(\ref{accuracy}). 
When this criterion is respected and hence the parameters of the polynomial are fixed,
the statistical error on $\langle {\cal O} \rangle$ will only be a function
of $N_{\rm corr}$. In the following we will see which values of $N_{\rm corr}$ 
are sufficient to keep the error on $\langle {\cal O}\rangle$ small.
 
A most important quantity in determining the cost of a simulation of
a given algorithm is
the autocorrelation time. Since we are using the correction factor
to render the algorithm exact,
all observables have to be computed as a ratio of $\langle {\cal O}W\rangle_P$ and
$\langle W\rangle_P$: hence it is not obvious how to define the autocorrelation
function of ${\cal O}$, in terms of which the integrated autocorrelation time
$\tau_{\rm int}({\cal O})$ is usually defined.
We ``define'' the integrated autocorrelation time for a given observable ${\cal O}$
by means of the expression which can be derived in the ordinary case,
when no reweighing occurs:
\be \label{tau_P}
 \tau_{\rm int}({\cal O}) =
\frac{1}{2}\left( \frac{ \sigma({\cal O}) }{ \sigma^{\rm naive}({\cal O}) } \right)^2 \;
\ee
where $\sigma^{\rm naive}({\cal O})$ denotes the naive and $\sigma({\cal O})$ the true error
on the observable ${\cal O}$.

In order to obtain a reliable estimate of the true error on
${\cal O}$ in all of our tests, which are discussed below, 
we have used a single elimination jack-knife procedure.
The jack-knife procedure has been then combined with a binning analysis
by blocking the data into blocks of length $L_{\rm block}$.
Our error analysis follows closely the discussion in \cite{jansom}
(see section 5.2 there).

We have run $K$ replica in parallel and determined the true error
in two ways: In the first approach we average on each replicum separately.
Since the averaged data are statistically independent, we can estimate the
true error by looking at the naive dispersion of them with respect to their 
arithmetic average. The {\em relative} error on the error in this case can 
be estimated as $(2K)^{-1/2}$.
In the second approach, we divide the sample into blocks of size
$L_{\rm block}$, so that the total number of blocks is given by
$N_{\rm block}=KN_{\rm traj}/L_{\rm block}$,
where $N_{\rm traj}$ is the number of trajectories obtained per replicum.
Of course, $L_{\rm block}$ is
to be constrained by the requirement that data coming from different
replica never appear in the same block.
The error can then be computed as a function of the block length
$L_{\rm block}$. For a large enough block length, a plateau behaviour sets
in from which we then determine the true error.
The {\em relative} error on the error in this procedure can be
estimated as $(2N_{\rm block})^{-1/2}$.

Even when we have determined the true error on an observable as discussed above,
the definition of the naive error on $\langle {\cal O} \rangle$, and consequently
the autocorrelation time $\tau_{\rm int}({\cal O}) $, is not obvious, again due
to the occurrence of the reweighing factor $W$ in the definition of
$\langle {\cal O} \rangle$. A possible definition of the naive error on
$\langle {\cal O} \rangle$ is given  by the single elimination
jack-knife error for a block length of $L_{\rm block}=1$.
We remark however that the
variance of ${\cal O}$:
\be \label{apriori_var}
{\cal V}({\cal O}) = \langle {\cal O}^2 \rangle -  \langle {\cal O} \rangle^2
                   = \frac{\langle {\cal O}^2 W \rangle_P}{\langle W \rangle_P}
                   -  \frac{\langle {\cal O}W \rangle_P^2}{\langle W \rangle_P^2}
\ee
is an observable itself and should hence be independent of a particular
algorithm used to compute it.
The above observation suggests another definition of the naive error
on $\langle {\cal O} \rangle$, which is the one adopted in our studies of the PHMC
algorithm:
\be \label{sigma_naive}
\sigma^{\rm naive}({\cal O}) = \left[ (N -1)^{-1} {\cal V}({\cal O}) \right]^{1/2} \; .
\ee
with $N=KN_{\rm traj}$.
Note that only for $W=1$ both definitions of the naive error have to agree. 

\subsection{Tuning of the PHMC algorithm}

\begin{figure}
\vspace{0.0cm}
\begin{center}
\psfig{file=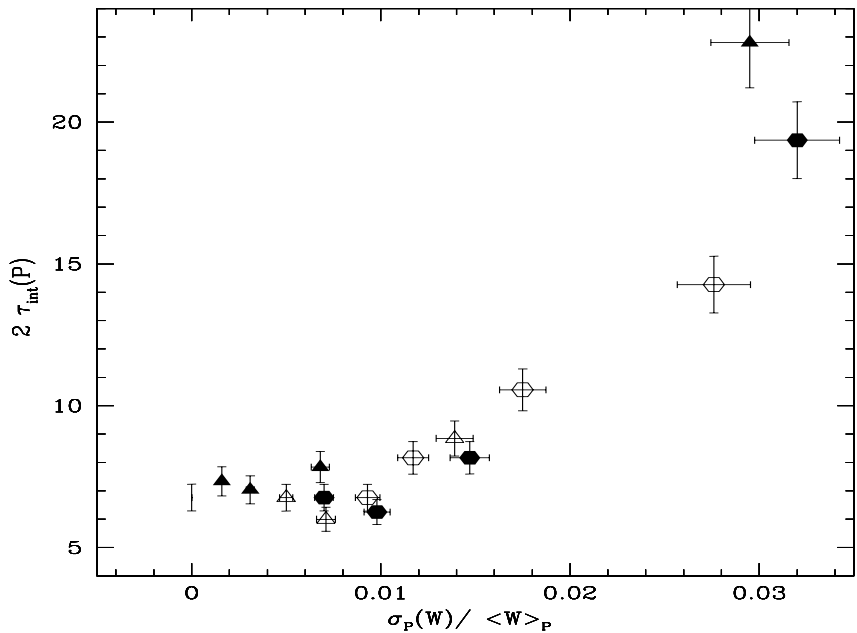, %
width=12cm,height=9cm}
\end{center}
\caption{ \label{fig:tau_int_P} The integrated autocorrelation time for the plaquette, $2\tau_{\rm int}(P)$,
versus $\sigma_P(W)/ \langle W \rangle_P$, for several values of $\epsilon$:
$\epsilon = 0.046$ (empty hexagones), $\epsilon = 0.036$ (filled hexagones),
$\epsilon = 0.026$ (empty triangles), $\epsilon = 0.016$ (filled triangles).
For each value of $\epsilon$ four values of $n$ ($8$, $12$, $16$, $20$) are
considered: the smaller is $n$, the larger is the corresponding value of
$\sigma_P(W)/ \langle W \rangle_P$ in the plot.
}
\end{figure}

In order to investigate on a quantitative level the tuning problem for $n$, $\epsilon$ and 
$N_{\rm corr}$, we have run the PHMC algorithm on a $4^4$
lattice with Schr\"odinger functional boundary conditions \cite{su3paper,sint} 
for a number of choices
of $n$ and $\epsilon$.  
To be more specific, we have set $c_t(g_0)=1$ and $\theta=\pi/5$. 
At the boundary at time $t=0$ the gauge fields were set to classical fields
denoted as point ``A'' in \cite{su3paper}. Finally, the gauge fields at time
$t=T$ were set to be identical to the one at $t=0$.
We remark that since we have chosen the gauge fields
to be identical at both time boundaries, we do {\em not} have exactly
the same boundary conditions as the ones in \cite{su3paper}.
The simulation parameters were chosen to be             
$c_{\rm sw}=0$, $\beta=6.4$ and $\kappa =0.15$. Although in this situation 
the average condition
number for $\hat{Q}^2$ was only about $60$ we still find a significant dependence
of the simulation cost on the algorithm parameters such that a sensible
tuning can be performed. 
For each choice of $n$ and $\epsilon$ about $20000$
trajectories were generated using a step size of 
$\delta\tau = 0.25$ and the number of molecular dynamics steps
$N_{\rm md} = 4$. These parameters were chosen to yield  
acceptance rates of about $80\%$. The values of $N_{\rm corr}$  
were varied from 1 to 4 or 1 to 10 in these simulations. 
These values for $N_{\rm corr}$ turned out to be sufficient 
to look for an optimal value minimizing the computational cost.
We will consider here mainly two observables,
the plaquette and the lowest eigenvalue of $\hat{Q}^2$, 
denoted by $P$ and $\lambda_{\rm min}$,
respectively. 
We mention that we monitored also
the largest eigenvalue of $\hat{Q}^2$ and the
reweighing factor itself. Within statistical errors
the average values for all the considered observables agree among all our
simulations with the PHMC algorithm 
and with the corresponding results obtained from the HMC algorithm. 

\begin{figure}
\vspace{0.0cm}
\begin{center}
\psfig{file=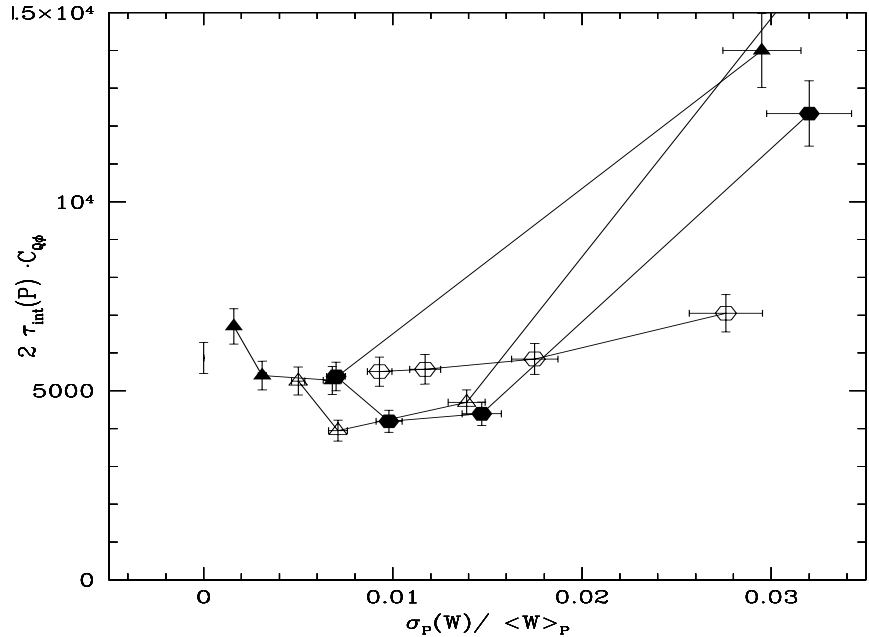, %
width=12cm,height=9cm}
\caption{ \label{fig:cost_meas_P} The cost of obtaining a statistically independent measurement of the plaquette,
$2 \tau_{\rm int}(P) \cdot C_{Q\phi}$ is plotted versus $\sigma_P(W)/ \langle W \rangle_P$,
using the same symbols as in Fig.\ref{fig:tau_int_P}. 
}
\end{center}
\end{figure}

We start by showing the
integrated autocorrelation time of the plaquette observable in
Fig.~\ref{fig:tau_int_P}. 
As discussed above we expect the cost of a simulation using the PHMC algorithm to 
depend strongly
on the relative fluctuation of the correction factor. We therefore plot the
integrated autocorrelation time as a function of 
$\sigma_P(W)/ \langle W \rangle_P$. Here the subscript $P$ (not to be confused
with the symbol for the plaquette) reminds that the mean value and the standard 
deviation for the correction factor $W$ do not involve, of course, any reweighing and
\be \label{ave_sigma_P}
\sigma_P(W) = \left[ (N_{block} - 1)^{-1} ( \langle W^2 \rangle_P
 - \langle W \rangle_P^2 ) \right]^{1/2} \; .
\ee
For the figure we have chosen four values of $\epsilon$ ($0.046$, $0.036$, $0.026$,
$0.016$) and $n$ ($8$, $12$, $16$, $20$). 
The smaller $n$ is, the larger is the corresponding value of
$\sigma_P(W)/ \langle W \rangle_P$ in the plot. For each choice
of $n$ and $\epsilon$, we took the value of $N_{\rm corr}$ that turns out to
minimize the quantity $2\tau_{int}(P) C_{Q\Phi}$ (see below).
The point at $\sigma_P(W)/ \langle W \rangle_P=0$
belongs to the integrated autocorrelation time as obtained from the HMC algorithm.
It is clearly seen that when increasing $\sigma_P(W)/ \langle W \rangle_P$ the 
integrated autocorrelation time assumes large value. For 
$\sigma_P(W)/ \langle W \rangle_P<0.01$, the dependence of the 
autocorrelation time becomes weak and no preferred choice of $n$ and $\epsilon$
can be given.

In Fig.~\ref{fig:cost_meas_P} we show the total cost by
computing $2\tau_{\rm int}(P) C_{Q\Phi}$, with the cost factor $C_{Q\Phi}$ 
given in eq.(\ref{PHMC_cost}), again taking for each $n$ and $\epsilon$
the value of $N_{\rm corr}$ that minimizes $2\tau_{\rm int}(P) C_{Q\Phi}$ itself.
Here we find that for $\sigma_P(W)/ \langle W \rangle_P\approx 0.01$ one reaches 
the minimal cost of the algorithm. 
We also give, at $\sigma_P(W)/ \langle W \rangle_P = 0$,
the cost of a corresponding HMC simulation. As already mentioned in \cite{frezzi}
the cost obtained from the PHMC algorithm is significantly lower. 
On the other hand, for $\sigma_P(W)/ \langle W \rangle_P > 0.01$
the cost from the PHMC algorithm increases, which is a direct consequence of
the increase of the autocorrelation time observed in Fig.~\ref{fig:tau_int_P}. 
For $\sigma_P(W)/ \langle W \rangle_P \ll 0.01$ we also find an increase of 
the cost of the simulation. 
This is a consequence of the
fact that more precise polynomial approximations have a higher computational
cost without giving any sensible reduction of the autocorrelation time $\tau_{\rm int}(P)$
and, correspondingly, of the statistical error $\sigma(P)$.
The corresponding results for $\lambda_{\rm min}$ look qualitatively similar, although
with a somewhat stronger dependence on $\sigma_P(W)/ \langle W \rangle_P$. 
Also in this case we find the optimal value of $\sigma_P(W)/ \langle W \rangle_P \approx 0.01$.

\begin{table*}[hbt]
\setlength{\tabcolsep}{1.5pc}
\vspace{2mm}
\begin{tabular*}{\textwidth}{@{}l@{\extracolsep{\fill}}llll}
\hline
$L^3 \times T \;\;$ & Algor.($N_{\rm corr}$) & $< P >$ & $
< \lambda_{\rm min}(\hat{Q}^2) >$ \\
\hline \hline
 $4^4$ & HMC & $0.66179(5)[12]$
          & $0.01582(3)[8]$  \\
\hline
 $4^4$ & PHMC(4)  & $0.66186(5)[12]$
       & $0.01582(3)[8]$  \\
       & PHMC(3)  & $0.66185(5)[12]$
       & $0.01583(3)[8]$  \\
       & PHMC(2)  & $0.66185(5)[12]$
       & $0.01583(3)[8]$  \\
 $\rightarrow $
       & PHMC(1)  & $0.66185(5)[12]$
       & $0.01583(3)[8]$  \\
       & PHMC(0)  & $0.66221(5)[12]$
       & $0.01451(3)[8]$  \\
\hline
 $4^4$ & PHMC(10) & $0.66188(5)[18]$
       & $0.01588(3)[16]$  \\
       & PHMC(9)  & $0.66188(5)[18]$
       & $0.01584(3)[16]$  \\
       & PHMC(7)  & $0.66198(5)[19]$
       & $0.01586(3)[17]$  \\
       & PHMC(5)  & $0.66198(5)[20]$
       & $0.01581(3)[17]$  \\
 $\rightarrow $
       & PHMC(4)  & $0.66201(5)[22]$
       & $0.01584(03)[18]$  \\
       & PHMC(3)  & $0.66213(5)[23]$
       & $0.01575(3)[19]$  \\
       & PHMC(2)  & $0.66215(5)[28]$
       & $0.01569(3)[22]$  \\
       & PHMC(1)  & $0.66218(5)[36]$
       & $0.01553(3)[24]$  \\
       & PHMC(0)  & $0.66272(5)[16]$
       & $0.01218(3)[8]$  \\
\hline \hline
\end{tabular*}
\caption{The behaviour of mean values and statistical errors for the plaquette and the
 lowest eigenvalue of $\hat{Q}^2$ as a function of $N_{\rm corr}$: data are obtained with the HMC 
 and the PHMC algorithms. For the latter we have considered the parameters 
 $n=8$, $\epsilon=0.036$ (data set with $N_{\rm corr}$ ranging from $1$ to $10$)
 and $n=16$, (data set with $N_{\rm corr}$ ranging from $1$ to $4$).
 The statistics has been $21000$ trajectories in all cases.
 We give in round brackets the naive error and in square brackets our
 estimate for the true error. 
 An arrow points towards the line where the value of $N_{\rm corr}$ turns out to be 
 basically optimal. 
 The case $N_{\rm corr}=0$ in the PHMC data refers to the results 
 obtained with no reweighing.}  
\label{tab:N_corr_dep44}
\end{table*}

After having identified the optimal values for $n$ and $\epsilon$, it is 
interesting to study the behaviour of the 
statistical errors as a function
of the values of $N_{\rm corr}$.  
To this end, we have chosen two different values of $n$ and $\epsilon$.
The first one, $n=16$ and $\epsilon=0.026$ corresponds to 
$\sigma_P(W)/ \langle W \rangle_P \approx 0.01$ and is therefore considered
to be close to the optimal value. 
The other choice is $n=8$ and $\epsilon=0.036$, which gives a value of
$\sigma_P(W)/ \langle W \rangle_P \simeq 0.032$ and is clearly far from being 
optimal. The mean values for the plaquette $P$ and eigenvalue $\lambda_{\rm min}$ as well as
the naive (round bracket) and true (square bracket) errors are given 
in table~\ref{tab:N_corr_dep44}. 
The value of $N_{\rm corr}=0$ corresponds to the case where no reweighing is 
performed, which is expected to yield systematically wrong results. 
For the non-optimal choice of $n$ and $\epsilon$ 
we observe a strong dependence of the statistical errors on $N_{\rm corr}$.
The lowest values of $\sigma$ are obtained only for $N_{\rm corr}=10$, i.e. the largest 
of 
the considered values of $N_{\rm corr}$. Even the values of $\sigma$ corresponding to
$N_{\rm corr}=10$ are still somewhat larger (especially for $\lambda_{\rm min}$) than both  
the statistical errors for $N_{\rm corr}=0$ and
from the HMC algorithm. 
This behaviour closely 
corresponds to what is expected from the above discussion about the statistical noise
induced by the reweighing procedure. On the other hand, when  
considering the optimal choice of $n$ and $\epsilon$, the statistical errors from
the PHMC algorithm do not show any visible dependence on $N_{\rm corr}$ and
basically coincide with the ones from the HMC algorithm. Finally we remark that
in all cases the mean values are consistent among themselves within the measured
statistical errors.
Moreover, the naive errors, defined according to 
eq.(\ref{sigma_naive}), are also consistent among all cases considered here. 

The behaviour of the error on the plaquette and $\lambda_{\rm min}$
was also tested on an $8^4$ lattice for 
parameter values $\beta
 =5.6$, $\kappa = 0.1585 \simeq \kappa_c$, $c_{\rm sw}=0$. The Schr\"odinger
functional boundary conditions that we adopted were chosen to be the same as for
the $4^4$ lattice mentioned above. The only difference is that the boundary
improvement coefficient was set to its 1-loop value, i.e. 
$c_t(g_0)=1.0-0.089g_0^2$.   
The statistics in this case is 
$2700$ trajectories.  We refer to \cite{frezzi} for a more   
detailed information about the algorithmic parameters
and give our results in table~\ref{tab:N_corr_dep88}. 
We compare the results obtained using the PHMC algorithm 
(in the setup with $K=32$ replica) with
the ones obtained using the HMC algorithm.  
We performed also a control run for the PHMC algorithm on only 1 replicum running up to 
the same statistics of 2700 trajectories. This gave completely
consistent results, as it should, of course, and provided us with a further 
check of our estimate of the uncertainty on the true error given in
braces in table~\ref{tab:N_corr_dep88}. From table~\ref{tab:N_corr_dep88} 
we infer that in this case the practically optimal
value of $N_{\rm corr}$ appears to be $2$ or $3$ and is hence again reasonably small
when a good polynomial approximation is chosen. 


\begin{table*}[hbt]
\setlength{\tabcolsep}{1.5pc}
\vspace{2mm}
\begin{tabular*}{\textwidth}{@{}l@{\extracolsep{\fill}}llll}
\hline
$L^3 \times T \;\;$ & Algor.($N_{\rm corr}$) & $< P >$ & $
< \lambda_{\rm min}(\hat{Q}^2) >$ \\
\hline \hline
 $8^4$ & HMC           & $0.57251(04)[12]\{3\}$
       & $0.001310(10)[51]\{8\}$  \\
\hline
 $8^4$ & PHMC(4)  & $0.57253(5)[14]\{3\}$
       & $0.001318(10)[50]\{8\}$  \\
       & PHMC(3)  & $0.57248(5)[14]\{3\}$
       & $0.001318(10)[50]\{8\}$  \\
 $\rightarrow $
       & PHMC(2)  & $0.57249(5)[15]\{3\}$
       & $0.001328(10)[50]\{8\}$  \\
       & PHMC(1)  & $0.57260(5)[19]\{5\}$
       & $0.001310(10)[60]\{10\}$  \\
       & PHMC(0)  & $0.57272(5)[12]\{2\}$
       & $0.001141(10)[45]\{7\}$  \\
\hline \hline
\end{tabular*}
\caption{ The behaviour of mean values and statistical errors for the plaquette and the
 lowest eigenvalue of $\hat{Q}^2$ as a function of $N_{\rm corr}$ on a $8^4$ lattice.
 The notation is the same as in table~\ref{tab:N_corr_dep44}. The numbers in braces
 give our estimate 
 of the uncertainty on the true error.} 
\label{tab:N_corr_dep88}
\end{table*}

The results on the 
$8^4$ lattice were obtained by taking $n=48$ and $\epsilon=0.0026$, yielding
a relative fit error $\delta \simeq 0.013$. This value of $\delta$ is even slightly 
larger than the
relative fit error corresponding to the optimal choice of $n$ and $\epsilon$
on the lattice $4^4$, when the condition number of $Q^2$ was about 10 times 
smaller. This seems to indicate that, even if the statistical fluctuations
of the correction factor are expected to increase with the lattice volume
and the condition number of $Q^2$, it might be unnecessary to take polynomial
approximations more and more severe.                                                  
This might be explained by the observation that
in this case also autocorrelation times generally increase, leading to a larger
number of evaluations of the correction factor on statistically correlated 
gauge configurations; in some cases, moreover, the statistical fluctuations 
of physical observables increase, too. However we think that further and
much more time-consuming studies are needed to clarify the issue.

\subsection{Exceptional eigenvalues}

So far, we have discussed the PHMC algorithm for situations where no
exceptional eigenvalues of $Q^2$, i.e. those that are orders of
magnitude smaller than the average lowest eigenvalue, appear.
However, the PHMC algorithm is designed to allow especially for
the occurrence of gauge field configurations
carrying exceptionally small eigenvalues of $Q^2$.
In fact we expect the probability of generating such configurations with
the PHMC algorithm to be considerably
larger than the corresponding probability when using
the HMC algorithm or exact versions of the multiboson technique with accept/reject step.

\begin{figure}
\vspace{0.0cm}
\begin{center}
\psfig{file=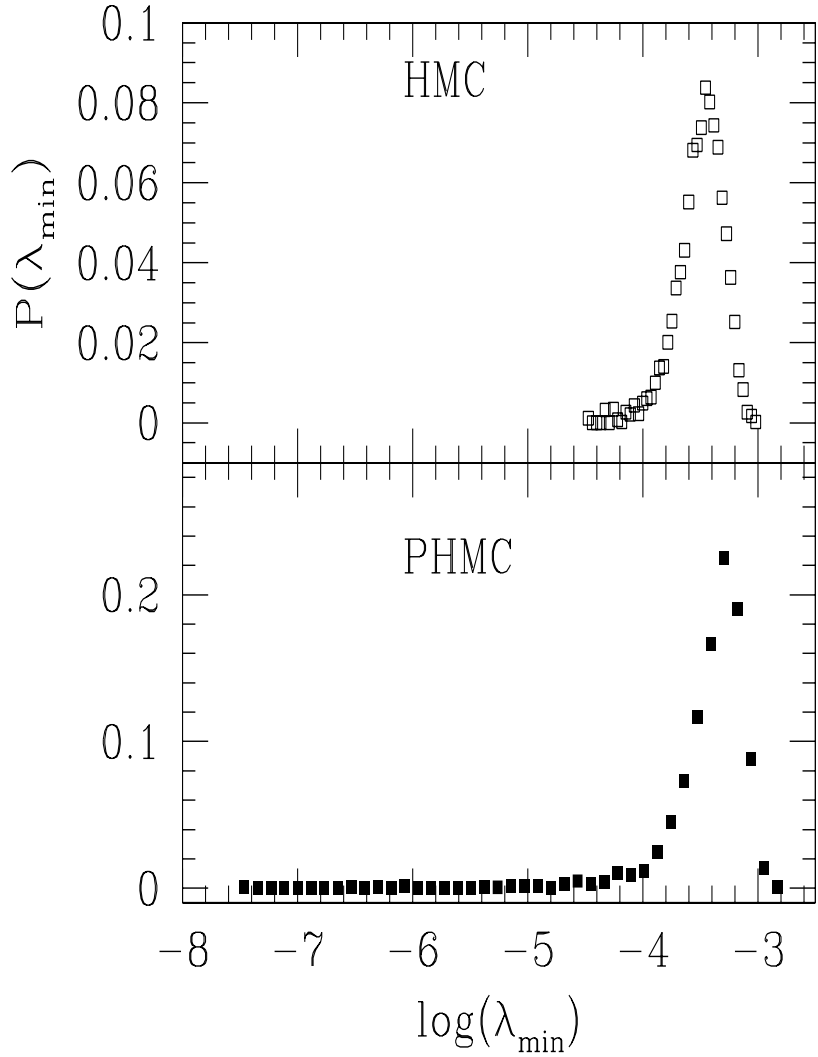,  width=12cm,height=16cm}
\end{center}
\caption{\label{fig:advertising} The distributions of the lowest eigenvalue 
$\lambda_{\rm min}$ of $\hat{Q}^2$ 
as obtained from 
the HMC and the PHMC algorithms. The quantity $P(\lambda_{\rm min})$ 
denotes the number of eigenvalues for a given bin, normalized 
by the total number of eigenvalues. Both runs have the same statistics.}
\end{figure}

This expectation is indeed confirmed in a real simulation, as can be seen
from Fig.~\ref{fig:advertising}. There we plot the distribution of the lowest
eigenvalue of $\hat{Q}^2$ 
as obtained from simulations with the HMC and the
PHMC algorithms. 
The parameters for the runs were chosen to be $\beta=5.4$, $\kappa =0.1379$ and 
$c_{\rm sw} = 1.7275$. The lattice size was  $8^3 \times 16$ and  Schr\"odinger
functional boundary conditions were adopted as specified in \cite{jansom}.  

\begin{figure}
\vspace{0.0cm}
\begin{center}
\psfig{file=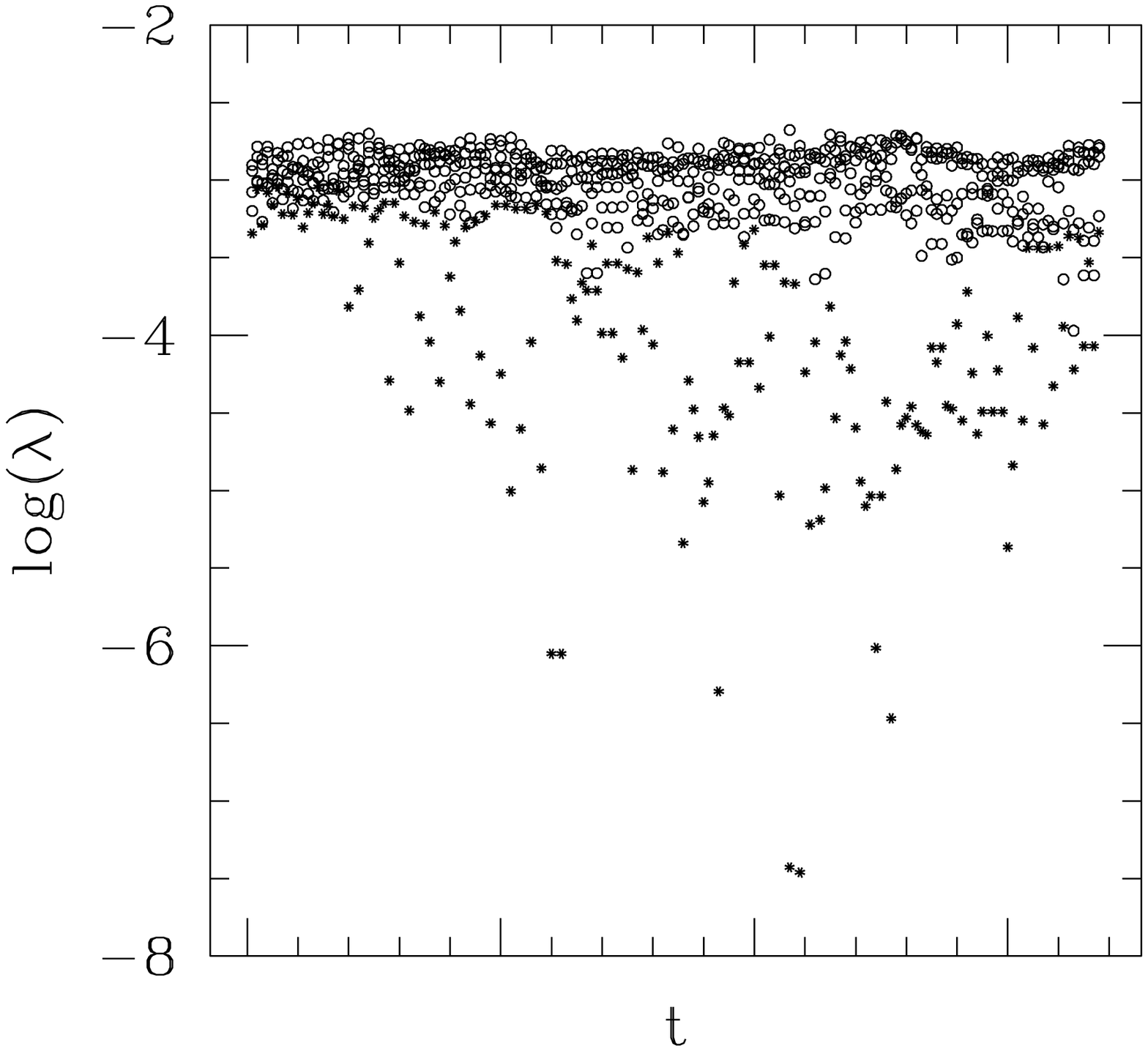,  width=12cm,height=12cm}
\end{center}
\caption{\label{fig:appetizer} The Monte Carlo time evolution of the five lowest eigenvalues
from a simulation using the PHMC algorithm. 
We denote by the stars the lowest eigenvalue and for the open
circles the remaining ones.
} 
\end{figure}

Clearly, the distribution obtained from the PHMC algorithm stretches to much
smaller values of $\lambda$. 
However, after reweighing with the correction factor, the average
lowest eigenvalue obtained in this case from the PHMC algorithm takes a value consistent
with the one obtained from the HMC algorithm.
In Fig.~\ref{fig:appetizer} we show the (Monte Carlo) time evolution
of the 10 lowest eigenvalues. We find that there is a band of eigenvalues
at a level roughly corresponding to the average lowest eigenvalue
and that only occasionally an {\em isolated} eigenvalue gets very small\footnote{
In some rare cases we have observed that the same happens for two or three eigenvalues.}.
This is exactly the situation anticipated in ref.~\cite{frezzi}.
As also discussed there, if $\lambda_{\rm min} \ll 1 $,
when computing the correction factor exactly on each gauge configuration,
i.e. taking $N_{\rm corr}=\infty$, $W= \det [Q^2 P_{n,\epsilon}(Q^2) ]$
turns out to be proportional to $\lambda_{\rm min}$.
Hence the correction factor serves the purpose of cancelling
divergences in certain quark Green functions. In the following
discussion we neglect the distinction between the operator $Q^2$,
which is certainly the relevant one for quark Green functions, and
some preconditioned form of it, which may be conveniently used in
the update and reweighing procedures. Indeed, doing so does not affect
our conclusions and keeps the discussion more general.

In practise the evaluation of the correction factor on gauge configurations carrying
exceptionally small eigenvalues may be problematic, since the badly
conditioned operator $Q^2 P_{n,\epsilon}(Q^2)$ has to be inverted
and $N_{\rm corr}$ is usually taken to be a finite (relatively small) number.

We see from eq.(\ref{W_eval_R}) that the quantity $[Q^2 P_{n,\epsilon}(Q^2)]^{-1}
\eta$ is needed for the evaluation of the reweighing factor as described above.
The inversion of $Q^2 P_{n,\epsilon}(Q^2)$ is performed by
using a CG algorithm, where suitable vectors are multiplied by $Q^2 P_{n,\epsilon}(Q^2)$
several times. As discussed in \cite{rootorder},
the multiplication of $Q^2 P_{n,\epsilon}(Q^2)$
is affected by rounding-error effects, which can be kept on a tolerable level
in normal situations.
However,
on gauge configurations carrying exceptionally small eigenvalues of $Q^2$, these
rounding-error effects might be significantly amplified, especially for the components
of $[Q^2 P_{n,\epsilon}(Q^2)]^{-1}\eta$ having non-vanishing projection on the low
lying mode eigenvectors.

In order to see the potential problems arising from taking a finite value of $N_{\rm corr}$,
let us introduce the eigenvalues $\lambda_j$ and eigenvectors $|\lambda_j \rangle$ of $Q^2$,
\be \label{Q2_eigen}
Q^2 |\lambda_j \rangle = \lambda_j |\lambda_j \rangle \; .
\ee
Then the correction factor
$W=\exp (w[\eta,U])$ of eq.(\ref{W_eval_R}) becomes
\bea    \label{W_spectral_form}
w[\eta,U] & = & \sum_j | \langle \lambda_j | \eta \rangle |^2 R_{n,\epsilon}(\lambda_j)
(1+ R_{n,\epsilon}(\lambda_j))^{-1}\; .  
\eea

Since $R_{n,\epsilon}(\lambda) \to -1$ as $\lambda \to 0$, all random fields $\eta$
that have a sizeable projection on the low--lying modes of $Q^2$ yield
very large negative values of $w$ and hence exponentially small values of $W$
are obtained.
The dominant contributions to the correction factor will
come
when the projection of the fields $\eta$ on the exceptional modes is almost zero.
In practise the correction factor is evaluated
stochastically by setting $N_{\rm corr}$ to a small value and
taking for the correction factor on a given gauge configuration $U$ the
following noisy estimate:
\be \label{eta_aver_W}
W[U;N_{\rm corr}] = N_{\rm corr}^{-1} \sum_{j=1}^{N_{\rm corr}} W[\eta_j,U] \; .
\ee
It is, of course, very unlikely that a field $\eta$ with almost
vanishing projection on the zero mode will be generated.
As a consequence, the noisy estimate of the reweighing factor obtained on a 
gauge configuration carrying low--lying modes of $Q^2$ is likely to be very imprecise.
On the other hand, an exact
evaluation of the reweighing factor of eq.(\ref{eta_aver_W}) with
$N_{\rm corr} = \infty$ will give that $W \propto \lambda_{\rm min}$ as
desired. 
 
The above discussion makes it clear that
the computation of the correction factor, as explained above for
``normal'' situations, should be generalized to deal with the case where
exceptional eigenvalues occur. To this end we introduce another
infrared cut-off parameter $\tilde{\epsilon} \ll \epsilon$ and write the partition
function as
\bea \label{phmc2partition}
{\cal Z} &
 = & \int {\cal D}U{\cal D}\phd{\cal D}\phi
{\cal D}\eta^\dagger{\cal D}\eta
  \; W_{B} W_{IR} \; e^{- (S_g + S_P + S_\eta ) } \nn \\
  S_P & = & S_P[U,\phi]=\phd P_{n,\epsilon}(Q^2[U]) \phi \nn \\
  S_\eta & = & \eta^\dagger\eta
\eea
where now the original correction factor $W$ is split into two parts,
an ``infrared'' part,
\be  \label{W_IR_eval}
W_{IR} =  \; \prod_{\lambda_j \le \tilde{\epsilon}} [ 1 + R_{n,\epsilon}(\lambda_j) ] \;
\ee
and a ``bulk'' part
\be \label{W_B_eval}
W_B[\eta,U]  = \exp \left\{\eta_{\perp}^\dagger [ R_{n,\epsilon} \cdot
               (Q^2\cdot P_{n,\epsilon})^{-1}] \eta_{\perp} \right\} \; ,
\ee
where
\be \label{eta_perp}
| \eta_{\perp} \rangle = | \eta \rangle - \sum_j \theta (\tilde{\epsilon} - \lambda_j)
| \lambda_j \rangle \langle \lambda_j | \eta \rangle \; .
\ee
The infrared part of the correction factor
$W_{IR}$ is very much in the spirit of ref.\cite{bunk}, where also the
underlying assumption was taken that only a few isolated small eigenvalues
occur.
Exact observables are now computed through
\be \label{N_corr_ave}
\langle {\cal O} \rangle = \langle W_B W_{IR} \rangle_P^{-1}
\langle {\cal O} W_B W_{IR} \rangle_P  \label{true_ave_2} \; .
\ee 
In order to guarantee the exactness of the simulation algorithm,
$\tilde{\epsilon}$ has to be {\em fixed} in a given simulation.
We give in appendix A a derivation of eq.(\ref{W_B_eval}) which
explicitly shows how the splitting of the original correction
factor $W$ into $W_B$ and $W_{IR}$ is fully determined {\em a priori}
by the choice of $\tilde{\epsilon}$.
Of course, when no eigenvalues smaller than $\tilde{\epsilon}$
occur, $W_{IR}=1$ and it has not to be computed.
In the case that such eigenvalues occur,
the two correction factors $W_B$ and $W_{IR}$ can be computed by
evaluating all eigenvalues $\lambda_j \le \tilde{\epsilon}$ and the
corresponding eigenvectors.

Obviously,
$W_B$ in eq.(\ref{W_B_eval}) receives no contribution
from the low-lying modes ($\lambda_j \le \tilde{\epsilon}$) of $Q^2$. This property
of $W_B$ justifies the expectation that a noisy reweighing with $W_B$ in
eq.(\ref{true_ave_2}) can be performed by choosing a small, finite value for
the number of $\eta$-field ``updates'' $N_B$.
From this point of view, the statistical noise induced
by reweighing with $W_B$ is expected to
be quantitatively very similar to the one induced
in ``normal'' situations by the reweighing factor $W$,
eq.(\ref{W_def}) and the
values for $N_B$ should in practise be similar to the values usually
chosen for $N_{\rm corr}$.

We have tested the modified correction factor in practise by taking a gauge
configuration carrying a mode of $Q^2$ with an exceptionally low eigenvalue
(about $3 \cdot 10^{6} $ times smaller than the highest one). 
Indeed, the estimate of $ \det [Q^2 P_{n,\epsilon}(Q^2) ]$  
obtained from the original reweighing factor $W$ is very imprecise
and converges very slowly to the correct value 
when increasing $N_{\rm corr}$. 
As a consequence one finds a large variance of $W$ as a function of $\eta$. 
On the other hand, the improved estimate of $W[U]$ given by
$W_{IR}[U] W_B[U;N_{B}]$, where 
\be \label{eta_aver_W_B}
W_B[U;N_B] = N_B^{-1} \sum_{j=1}^{N_B} W_B[\eta_j,U] 
\ee
and $\tilde{\epsilon} = \epsilon/10$, is much less noisy already for pretty small
values of $N_B$. In fact, the fluctuations are fully analogous to the case when no
exceptional eigenvalues are present.
More quantitative information on this point
will be given in a forthcoming publication \cite{phmc_perf}.

We remark that the problem of
inverting the operator $Q^2 P_{n,\epsilon}(Q^2)$ in the subspace orthogonal
to the one spanned by the low--lying modes of $Q^2$ is always well conditioned, even
in the presence of an exact zero mode.  The evaluation of $\eta_{\perp}$ can be done
by computing all the eigenvectors of $Q^2$ corresponding
to eigenvalues smaller than $\tilde{\epsilon}$.
Since, as shown in Fig.1, there are usually only a few isolated 
eigenvalues below
$\tilde{\epsilon}$, these eigenvalues and the corresponding eigenvectors can
be calculated reliably by using the techniques described in \cite{eigalgos}.

The level of precision needed in computing the low-lying
eigenvalues and the corresponding eigenvectors
of $Q^2$ is determined by requiring that the uncertainties in
$W_{IR}$ and $W_B$, induced by the uncertainties on these eigenvalues and eigenvectors,
be negligible (with respect to the statistical fluctuations
of $W_{IR}$ and $W_B$). 
Using the fact
that $R_{n,\epsilon}(\lambda) \simeq -1 + c \lambda$,
with $c = O(\epsilon^{-1})$, 
the relative uncertainty on each factor in the product of
eq.(\ref{W_IR_eval}) can be estimated (for $\lambda \ll 1$) as  
\be
[1+R_{n,\epsilon}(\lambda)]^{-1}  \delta [1+R_{n,\epsilon}(\lambda)] \simeq
\lambda^{-1} \delta \lambda
\nonumber \; ,
\ee
where $\delta \lambda$ denotes the uncertainty on a given eigenvalue $\lambda$. On the
other hand, the uncertainties in the determination of the eigenvectors corresponding to
eigenvalues smaller than $\tilde{\epsilon}$ directly affect the evaluation of
$| \eta_{\perp} \rangle$, eq.(\ref{eta_perp}), and hence $W_B$, eq.(\ref{W_B_eval}).

The discussion above makes very straightforward the software implementation of the
proposed reweighing procedure: once a value for $\tilde{\epsilon}$ has been set, the program
has just to check on each generated gauge configuration whether the lowest eigenvalue
of $Q^2$ is smaller than $\tilde{\epsilon}$. Only in the affirmative case, of course,
does the correction factor $W_BW_{IR}$ actually differ from the usual one and
an evaluation of $W_{IR}$ is required. 

It might be observed that, if the PHMC update ever generates gauge configurations with a
{\em significant} fraction of the modes of $Q^2$ belonging to very low eigenvalues, the above
described reweighing procedure becomes computationally very expensive. This is certainly
true, but in that case troubles are also expected in the evaluation of quark propagators by
ordinary CG--like inverters. Indeed, it is likely that in such a situation
a relatively precise knowledge of all the low--lying modes
is needed even for evaluating the fermion Green functions.
This can be done by splitting each quark propagator into two parts:
the first part, to which only the low-lying modes contribute, should be expressed
in terms of the known eigenvalues and eigenvectors;
the second part, to which no low-lying modes contribute, should
be evaluated by inverting the operator $Q$ in the subspace orthogonal
to the one spanned by the low--lying modes of $Q^2$.

%% file: section4_v2
\section{Rounding errors in the PHMC algorithm}

Rounding errors may become in principle a problem for all simulation
algorithms. Each algorithm is designed to produce field configurations
according to a probability distribution, related to the
Boltzmann factor of a given Euclidean field theory. 
The danger is that 
when 
implementing a code for some finite precision
computer, rounding errors 
may render 
the probability distribution of the actually produced field
configurations somewhat different from the desired one.
In particular, when using molecular dynamics kind of algorithms 
like the HMC or the PHMC algorithms, the equations of
motion, integrated numerically by a symplectic integrator,
lack in principle the reversibility condition, resulting in an inexact algorithm.
It is still an open question for what situations this systematic error of the
molecular dynamics kinds of algorithm will become important in practise. 

The problem of rounding errors ought to be studied especially for the PHMC
algorithm. As the discussion in section 2.2 showed, for an
efficient computation of $\delta S_P$ in the PHMC algorithm, 
the product representation
of the polynomial $P_{n,\epsilon}$ should be used. 
However, the stability of the numerical construction of the polynomial in
the product representation depends strongly on the ordering of the 
monomial factors in eq.~(\ref{pol_prod_q}). Particularly ``bad'' orderings easily lead
to substantial precision losses or even numerical overflow. 
As demonstrated in \cite{rootorder} (see also \cite{mon97}), there exist, fortunately, 
orderings of the monomial factors (or equivalently the roots of
the polynomial) such that rounding errors can be kept on a perfectly
tolerable level. 

Still, the rounding errors appearing for a particular ordering scheme applied 
in a given situation should be monitored carefully. In the generation 
of the pseudo-fermion fields $\phi$ one could in principle resort to 
numerically stable recursion relations. However, as discussed above, it is very
easy to monitor the rounding errors in this case by evaluating the difference
of eq.(\ref{check_phi}). 
In the evaluation of the correction factor, on the other hand, 
it is in general advisable to use a recursion relation \cite{foxy,numrec}  
in order to obtain
a sufficiently precise result.
In the following we will first discuss the rounding-error effects 
appearing in the construction of $Q^2 P_{n,\epsilon}(Q^2)$: this is the
polynomial of highest degree among the ones which occur in the PHMC
update and all the different polynomials of lower degree are numerically
constructed by using the same ordering of monomial factors. We then turn 
to a discussion of the magnitude of reversibility violations.

\subsection{Rounding errors from the product representation}

As shown in \cite{rootorder}, the Clenshaw recursion relation provides
a very stable and precise way to evaluate the polynomial 
$Q^2 P_{n,\epsilon}(Q^2) = 1 + R_{n,\epsilon}(Q^2)$, even
when 32 bit precision is employed everywhere, but in internal
products and other sums over the whole lattice.
This gives us the possibility of evaluating the size of the rounding errors
when the polynomial $Q^2 P_{n,\epsilon}(Q^2)$ is constructed in its 
product representation, eq.~(\ref{pol_prod_q}).
Following \cite{rootorder} we consider the vector 
\be \label{endvector}
 \Phi_{\rm order} \equiv \hat{Q}\sqrt{c_{n}} (\hat{Q}-r_{n}) \cdot \dots
  \cdot \sqrt{c_1} (\hat{Q}-r_1) \hat{Q}R_G \nonumber \\
\ee
where we have taken the preconditioned matrix $\hat{Q}$ as it was used in all
our numerical tests. 
The label ``order'' can stand for a particular  monomial ordering scheme. 
In the following we will only discuss the bit reversal and Montvay's schemes, 
which were found in \cite{rootorder} to be the most precise.
We refer again to \cite{rootorder} for a definition of the
ordering schemes employed here. 
Given the numerical stability of the Clenshaw recursion, a good 
measure of rounding errors, when only 32 bit precision is employed,
is the quantity 
\be \label{etahat}
\Delta_{\rm order} = \frac{1}{\sqrt{N}} \| \Phi_{\rm order} - \Phi_{\rm Clenshaw}\|\; .
\ee
In table~\ref{tab:eta_su3} we give the results for $\Delta_{\rm order}$ 
for the bit reversal and Montvay's
ordering schemes. All results have been obtained on an $8^3\times 16$ lattice with
Schr\"odinger functional boundary conditions as used for the 
computation for the ${\rm O}(a)$ improved action \cite{jansom}. 
The parameters of the runs were $\beta =  6.8$, $\kappa =0.1343$
and $c_{\rm sw} = 1.42511$.

\begin{table*}[hbt]
\begin{center}
\caption{The quantity $\Delta_{\rm order}$ eq.~(\ref{etahat}) is given for the 
bit reversal (BR) 
and Montvay's ordering schemes. 
}
\vspace{2mm}
\label{tab:eta_su3}

\begin{tabular}{lllll}
\hline
 $n$ & $\epsilon$ & $\delta$ & $\Delta_{\rm BR}$ & 
$\Delta_{\rm Montvay}$   \\
\hline \hline
 $16$   &  $0.0030$ & $0.310$   & $3.1(1) 10^{-6}$  & $3.0(1)\cdot 10^{-6}$  \\
 $32$   &  $0.0030$ & $0.054$   & $3.3(1) 10^{-6}$  & $3.0(1)\cdot 10^{-6}$  \\
 $64$   &  $0.0022$ & $0.0045$  & $4.5(1) 10^{-6}$  & $4.2(1)\cdot 10^{-6}$  \\
 $100$  &  $0.0022$ & $0.0002$  & $8.4(2) 10^{-6}$  & $6.4(2)\cdot 10^{-6}$  \\
 $100$  &  $0.0010$ & $0.0034$  & $9.0(2) 10^{-6}$  & $6.4(2)\cdot 10^{-6}$ \\
 $100$  &  $0.0005$ & $0.0218$  & $10.8(2) 10^{-6}$ & $7.8(2)\cdot 10^{-6}$ \\

\hline \hline
\end{tabular}
\end{center}
\end{table*}

We have chosen the constants $c_k$, eq.(\ref{pol_prod_q}), 
to be all identical. Choosing the $c_k$'s different from each other 
(while keeping fixed their product which guarantees the proper
normalization of $P_{n,\epsilon}$ ), changes the results 
in
table~\ref{tab:eta_su3} at most at the 10\% level.              
The results of table~\ref{tab:eta_su3} are
qualitatively very similar to the ones reported in \cite{rootorder}.
They show a growth of rounding errors
in the construction of the polynomial $P_{n,\epsilon}$ as $n$ and $\epsilon^{-1}$
increase 
(see the behaviour for $n=100$). 
However, 
the magnitude of rounding errors for the cases considered in 
table~\ref{tab:eta_su3} are perfectly tolerable. 
In particular, no evidence for numerical instabilities or large
rounding error effects has been observed.
Since all our simulations are performed using either the bit reversal
or Montvay's ordering schemes for a range of values of $n$ and $\epsilon$
covered by the ones
given in table~\ref{tab:eta_su3}, we conclude that our numerical simulations
are safe against rounding errors coming from the use of the product
representation for the polynomial $P_{n,\epsilon}$.

\subsection{Reversibility violations}

For the purposes of this section, the evolution of the gauge field in 
the molecular dynamics part of the PHMC algorithm can
be summarized as follows: 
some initial field configuration of the gauge fields $U_{x,\mu}$ and their
conjugate momenta $\pi_{x,\mu}$, $\left\{U_{\rm in},\pi_{\rm in}\right\}$,
is evolved from a fictitious Monte Carlo time $t=0$ to
the final configuration $\left\{U_{\rm end},\pi_{\rm end}\right\}$ at
$t=T$, with $T$ usually set to $T=1$ in production runs. 
This evolution is determined by the ``equations of motion", derived
from a Hamiltonian $H = \frac{1}{2}\sum_{x,\mu}\pi_{x,\mu}^2 + S$, where S is the total
action. 
At $t=T$, the
configuration $\left\{U_{\rm end},\pi_{\rm end}\right\}$ is subject
to an accept/reject step using the values of the Hamiltonians
$H_{\rm in}$ and $H_{\rm end}$, as measured
on the initial and final configurations, respectively.

We recall that in evolving the gauge field configuration 
in the Monte Carlo time
a great flexibility is allowed. The imposed restrictions are  
--from a practical point of
view-- that the acceptance rate determined by
$H_{\rm end} - H_{\rm in}$ should be reasonably large, about
$80\%$, and --from a principal point of view-- that the evolution in the
Monte Carlo time ought to be reversible in order to
guarantee detailed balance and consequently the correct importance
sampling.

The method of choice for the Monte Carlo time evolution is
to evolve the system with the equations of motion using a leap-frog
integrator. It was found,
in particular when machines with only 32-bit precision arithmetic are used,
that due to rounding errors, violations of the reversibility condition
are encountered. What is
worse, it appears that the equations of motion correspond to those
of a classical chaotic system with a positive Liapunov
exponent \cite{kramers,wuppertal,brower,ivan,revus}.
As a consequence, rounding error effects are exponentially
amplified along the integration of the equations of motion.

Using a leap-frog integrator --in particular on an APE machine 
with 32-bit arithmetic as
in this work-- needs therefore an estimate of violations of
reversibility. As it was discussed at length in \cite{rootorder},
in the PHMC algorithm some orderings of the monomial factors
in the product representation can lead to large rounding-errors effects 
with a possible strong influence on reversibility violations.
We therefore checked the magnitude of the 
reversibility violations when using the subpolynomial, the bit reversal and
Montvay's ordering schemes as described in \cite{rootorder}. 
These tests were performed with the same parameters as in Section~4.1. 
In particular, the polynomial parameters were chosen to be 
$n=64$ and $\epsilon = 0.0022$. 

To measure the reversibility violations, we simply started from 
the final configuration 
$\left\{U_{\rm end},\pi_{\rm end}\right\}$, reversed the sign of the step size 
$dt$ and integrated back to reach the reversed configuration 
$\left\{U_{\rm rev},\pi_{\rm rev}\right\}$. In all our tests we used 
the higher order leap-frog integrator as suggested in \cite{sexy}
(i.e. eq.(6.7) of that reference with $n=4$). 
Our 
step size was chosen to be  
$dt=0.05$ for both the forward and the backward integration and the value of 
the trajectory length was $T=0.75$. 
On the initial and the reversed configurations we measured
the corresponding Hamiltonians $H_{\rm in}$ and $H_{\rm rev}$ and 
the plaquettes $P_{\rm in}$ and $P_{\rm rev}$ averaged over the gauge configuration.   
The difference of these quantities, $dH$, $dP$ and the norm difference
$dU$ of the gauge links 
\begin{eqnarray} \label{dstuff}
\| dU\| ^2 & = & \|U_{\rm in} - U_{\rm rev} \|^2 = 
\frac{1}{36V}\sum_{x,\mu,\alpha,\beta} |
 U^{x,\mu,\alpha,\beta}_{\rm in} - U^{x,\mu,\alpha,\beta}_{\rm rev}| ^2 \nonumber \\ 
dH & = & \left| H_{\rm in} - H_{\rm rev} \right| \nonumber \\ 
dP & = & \left| P_{\rm in} - P_{\rm rev} \right| 
\end{eqnarray}
serve as our quantitative measure of the reversibility violations. 
In eq.(\ref{dstuff}) the sum extends over the lattice points, the 4 forward directions
and the colour indices.

\begin{table*}[hbt]
\caption{Reversibility violations for the PHMC and HMC algorithms,
         comparing different root orderings for the PHMC algorithm,
         subpolynomial (SP), bit reversal (BR) and Montvay's 
         ordering scheme. BR$^{*}$ indicates
         that the roots are calculated in 64-bit arithmetic.}
\vspace{2mm}
\label{tablerevers1}
\begin{center}
\begin{tabular}{lll}
\hline
Scheme & \makebox[1.5cm][r]{$\langle \| dU \|\rangle$}& $\langle dH\rangle $ \\ 
\hline \hline
 SP      & $9.45(1)\cdot 10^{-6} $ & $2.1(2)\cdot 10^{-2}$     \\
 BR      & $1.293(1)\cdot 10^{-6} $ & $4.0(9)\cdot 10^{-3}$     \\
 BR$^{*}$& $1.292(1)\cdot 10^{-6} $ & $2.8(8)\cdot 10^{-3}$     \\
Montvay  & $1.277(1)\cdot 10^{-6} $ & $3.4(9)\cdot 10^{-3}$     \\
 HMC     & $6.7(2)\cdot 10^{-7}  $ & $1.4(6)\cdot 10^{-3}$     \\
\hline \hline
\end{tabular}
\end{center}
\end{table*}

Our results, averaged over 32 configurations
are given in table~\ref{tablerevers1} for the subpolynomial
(SP), the bit reversal (BR) and Montvay's ordering scheme. We compare with
the corresponding results from the HMC algorithm, 
using there the same number of steps
and an equal step size as used in the case of the PHMC algorithm. 
For the HMC algorithm, in the Conjugate Gradient solver
we have chosen a stopping criterion requiring that the squared norm
of the residual vector, normalized by the solution vector, be
less than $\epsilon_{\rm stop}^{\rm HMC} = 10^{-14}$.

One clearly sees that the subpolynomial scheme
gives substantially more reversibility violations than the 
one encountered in the HMC algorithm.
Within the errors, the size of the reversibility violations from the PHMC 
algorithm with the bit reversal and Montvay's scheme are 
of the same order as the ones from the HMC algorithm. We also considered the bit reversal
ordering in the case, denoted as BR$^{*}$, when the roots and the normalization 
factors are computed with 64-bit precision and then read in. Within the errors
we do not find any effect. In our tests we could find no difference
in the plaquette expectation value. 
We conclude that our results are not contaminated
from reversibility violation effects.  

As mentioned above, there is a lot of flexibility to perform the 
evolution of the gauge field configuration in Monte Carlo time in the 
molecular dynamics part of the HMC algorithm. 
The PHMC algorithm establishes an approximation of the exact 
evolution. The crucial advantage of the PHMC algorithm is, 
of course, that this approximation is fully controlled
and can be corrected for. 
Another possibility of approximating the Monte Carlo time
evolution is to just use a larger stopping criterion 
for the inverter of $\hat{Q}^2$. However, we think that the
reversibility violations, which are certainly present, 
may then become dangerous: due to rounding errors, when integrating the gauge fields 
backward in time, the inverter ``sees'' a different gauge
field configuration from the one during the forward integration. 
Therefore also the solution vectors will be different and when
the stopping criterion is relaxed, this difference is enhanced,
possibly leading to large reversibility violations.

This effect is amplified when one makes use of the 
``knowledge of the past'': 
the inverter is started with an initial guess, which is the
solution of the previous inversion. This reduces the number
of iterations in the inverter, since generally the movement of
the gauge fields through configuration space is smooth. The idea 
may also be iterated \cite{brower}. However, in this way 
potential rounding errors are amplified, since they are accumulated
in the solution vectors. 

A possible solution may be to choose always a constant starting vector.
Then reversibility violations only appear through the difference in the
gauge field configuration. We tested this possibility for our
implementation of the HMC algorithm and report our results in table 
\ref{tablerevers2}. A similar investigation has been performed
in \cite{ivan}. Here we have taken the same parameters
as used for table~\ref{tablerevers1}, averaging again over 32 configurations. 

\begin{table*}[hbt]
\caption{Comparison of reversibility violations in the HMC algorithm using 
a constant starting vector in the Conjugate Gradient solver.} 
\vspace{2mm}
\label{tablerevers2}
\begin{center}
\begin{tabular}{llll}
\hline
$\epsilon_{\rm stop}^{\rm HMC}$ & \makebox[1.5cm][r]{$\langle \| dU \| \rangle$}& $\langle dH\rangle $ 
         &   $\langle dP \rangle $  \\
\hline \hline
$1.0\cdot 10^{-14}$  & $6.58(1)\cdot 10^{-7} $ & $1.1(6)\cdot 10^{-3}$  & $--$    \\
$1.0\cdot 10^{-12}$  & $8.5(1)\cdot 10^{-7} $   & $1.8(7)\cdot 10^{-3}$  & $--$    \\
$1.0\cdot 10^{-10}$  & $2.6(4)\cdot 10^{-6} $ & $7.8(1.8)\cdot 10^{-3}$ & $6(3)\cdot 10^{-8}$  \\
$1.0\cdot 10^{-8}$   & $4.7(3)\cdot 10^{-5} $   & $2.0(3)\cdot 10^{-1}$  & $6(1)\cdot 10^{-7}$ \\
\hline \hline
\end{tabular}
\end{center}
\end{table*}

As can be seen, already for a stopping criterion of $10^{-10}$ the reversibility violations
are substantially larger than for the severe stopping criterion, showing even a difference
in the expectation value of the plaquette. We conclude that on machines with 32-bit precision
arithmetic the stopping criterion can not be relaxed too much. 
Since with a constant starting vector we loose the advantage of having
a reasonable first guess for the solution of the inverter, we 
prefer 
using a severe stopping criterion and a better initial guess over 
relaxing the stopping criterion and using a constant starting
vector.                 
Of course, the situation might
look different on machines with higher precision, where reversibility violations
are suppressed. 

We would like to point out a second effect relevant for reversibility violations.
When the stopping criterion is made large,
it might happen that during the backward integration the inverter stops one
iteration before or later than on the corresponding step in the forward 
integration. Since now the stopping criterion
is large, the solution vectors are very different, leading to large
reversibility violations. 
The only way to overcome this would be
to also keep the number of iterations constant. 
However, we feel that with this way of accelerating the algorithm
the convergence of the inverter is not very well controlled, but
we have not studied this situation in detail and we do not know
how a possible poor convergence may affect the acceptance of the
whole molecular dynamics trajectory.

%% file: section5_v2
\section{Memory requirements}


In this section we wish to discuss three possible ways of reducing memory
requirements. The first two ways (sections 5.1 and 5.2) result in some 
reasonably tolerable computational overhead. The last way (section 5.3)
was already shortly mentioned in \cite{frezzi} but it 
leads to a significant alteration of the dynamics.

Once again, we neglect in our discussion the technical
complications arising from the use of even--odd preconditioning, which
can however be treated as in any HMC--like algorithm. We recall here that
the pseudofermion fields $\phi$ and $\chi_k, \quad k=1,2,\dots $, which
will enter our discussion, are assigned to arrays defined on all lattice
sites. Indeed, even if only the odd--site components are needed in principle,
we have found it convenient to make use of the even--site components to store 
intermediate results in the construction of the operator $\hat{Q}$
(see eq.(35) of \cite{rootorder}), which connects second neighbour 
sites on the original lattice.

\subsection{PHMC update with $(Q^2-z_k)$ monomials}

It is, of course, possible to implement the PHMC algorithm by using
also the product representation 
\begin{equation} \label{prod_q2}
P_{n,\epsilon}(Q^2) = \prod_{k=1}^{n} [c_k(Q^2-z_k)] 
\end{equation}
with the roots $z_k$ given in eq.(\ref{roots_rk}). 
The variation of the action $S_P$ then becomes 
\be \label{PHMC_force_q2} 
\delta S_P =  \sum_{k=1}^{n/2} c_k
            \left[ \delta Q^2 \; \xi_{k-1} \otimes \xi_{n-k}^{\dagger}
            + \delta Q^2 \; \xi_{n-k} \otimes \xi_{k-1}^{\dagger} \right] \; ,
\ee
where the auxiliary pseudofermion fields $\xi_k$, for $k=1,\dots ,n-1$ are
given by
\be  \label{auxphi_q2_def}
\xi_k \equiv [c_k(Q^2-z_k)] \cdot [c_{k-1}(Q^2-z_{k-1})]
             \cdot  \dots \cdot [c_1(Q^2-z_1)] \phi \; .
\ee

Following the discussion in section 2.2, the evaluation of $\delta S_P$ in
eq.~(\ref{PHMC_force_q2}) implies a memory requirement of 
only $(n/2) + 2$ pseudofermion fields, which means a reduction of basically a factor 2. 

However, if one insists on using only $(n/2)+2$ pseudofermion fields, it
seems impossible to avoid an overhead on the computational cost.
In evaluating $\delta S_P$ in eq.(\ref{PHMC_force_q2}) one needs, analogously
to the case discussed in Section 2.2, $3n$ $Q\phi$ operations. 
There appear, however, 
additional $n/2$ $Q\phi$ operations for the following reason. 
When
storing only the fields $\xi_1, \dots , \xi_{n/2}$ before starting the computation
of $\delta S_P$, one is ``loosing'' the information about the vectors                   
$Q \xi_1, \dots , Q \xi_{n/2-1}$, which have already been calculated as intermediate steps
in the evaluation of the auxiliary fields $\xi_1, \dots , \xi_{n/2}$. 
This information is needed since 
$\delta Q^2 = \left[ \delta Q \right]Q + Q \left[ \delta Q \right]$.
A similar problem with the vectors $Q \xi_l $ for $l=n/2, \dots ,
n-1$ can be avoided by a prudent usage of memory space
associated with the vectors $\xi_l$. As a consequence, 
with respect
to the method described in Section 2.2, when using the product
representation of eq.(\ref{prod_q2}), the memory requirements are 
basically halved at the price
of an increase of the computational cost of $\delta S_P$, which can be
estimated to be about 15--20\%.

\subsection{Flexible trading between memory requirement and CPU time}

It is clear that in implementing the evaluation of $\delta S_P$, eq.(\ref{PHMC_force}),
one can trade off CPU time with memory space in different ways. We sketch here
the basic idea of a flexible method for compromising between memory and CPU
time, which we have found very convenient in practical simulations. For simplicity
we take the example of a polynomial, $P_{n,\epsilon}(Q^2)$, eq.(\ref{pol_prod_q}),
with $n=100$ and consider a non-optimized version of the method that we use in practice.
A fully general and very detailed technical discussion of this method and 
its performance is deferred to Appendix B.

A significant fraction of the memory is usually taken to store the gauge links, their 
conjugate momenta, some pseudofermion vectors, as needed
for the fermion matrix inversion, and the dynamical pseudofermion
field, $\phi $, extracted from a probability distribution 
$\propto \exp{[- \phd P_{n,\epsilon}(Q^2) \phi]}$.  
Let us imagine to divide the remaining storage space (assumed to be much less
than what is needed for storing $n=100$ pseudofermion vectors) into three sectors.
In this particular case, with $n=100$, the first and the second sector will contain 
$9$ and the third sector only $2$ pseudofermion vectors. It is clear that the
third sector can be used as working space for fermion matrix times vector multiplications,
where neither the initial vector nor the final one need to be stored elsewhere. 

We have already observed in Section 2.2 that the variation $\delta S_P$,
eq.(\ref{PHMC_force}), of the pseudofermion action $S_P$, is a sum of
$n$ terms. Each term depends only on two auxiliary fields, $\chi_{j-1}$,
$\chi_{2n-j}$ (and their complex conjugates), where $j$ is the index
over which the sum runs and the auxiliary vectors $\chi_k$ are defined
in eq.(\ref{auxphi_def}). For the evaluation of $\delta S_P$ one can
then proceed as follows.

In a preliminary step, starting from $\phi$, we construct the auxiliary vectors
$\chi_1$, $\chi_2$, \dots, $\chi_{89}$, $\chi_{90}$, and store {\em only} 9 vectors,
namely $\chi_{10}$, $\chi_{20}$, \dots, $\chi_{80}$, $\chi_{90}$, in the part
of the memory that we have indicated above as the first sector.

Then, in a first step, starting from the saved vector $\chi_{90}$,
we construct the auxiliary vectors $\chi_{91}$, $\chi_{92}$,
\dots, $\chi_{98}$, $\chi_{99}$ and store all of them in the second sector. We are
now in position to evaluate the first ten contributions to $\delta S_P$, namely
the ones corresponding to $j=100,99, \dots , 91$ in eq.(\ref{PHMC_force}). The point
is that fermion matrix times vector multiplications can be performed in such a way that
the third sector is employed to store {\em in turn} first $\chi_{100}$ and $\chi_{101}$, 
then $\chi_{101}$ and $\chi_{102}$, and so on, up to $\chi_{108}$ and $\chi_{109}$.

In the second step, starting from the saved vector $\chi_{80}$, 
we construct the auxiliary vectors $\chi_{81}$, $\chi_{82}$,
\dots, $\chi_{88}$, $\chi_{89}$ and store all of them in the second sector. We are
now in position to evaluate further ten contributions to $\delta S_P$, namely
the ones corresponding to $j=90,89, \dots , 81$ in eq.(\ref{PHMC_force}), making
use of the third sector to temporarily store the various pairs of auxiliary vectors
between $\chi_{110}$ and $\chi_{119}$, as explained above.

Proceeding in an analogous way, we can evaluate in ten steps all the
contributions to $\delta S_P$. Notice that in each of these steps, except the
first one, nine pseudofermion vectors, which had been computed and immediately overwritten
during the preliminary step mentioned above, are computed again. This leads to a
global computational cost, which is equivalent to about $390$ $Q\phi$ operations,
to be compared with the cost of about $300$ $Q\phi$ operations needed for the method
discussed in section 2.2 (for a single evaluation of $\delta S_P$). This increase
of the computational cost is just the price to be paid for evaluating $\delta S_P$
in the case $n=100$ by using only $20$ (instead of $100$) auxiliary pseudofermion
vectors. A similar result, with somewhat better compromise between memory and CPU
time, is found for any value of the degree $n$ of the PHMC polynomial when using 
the generalized version of this method which is described in Appendix B.

Let us conclude with a general remark about the well-known problem of 
large memory requirements, which is in principle common to all algorithms
for dynamical fermions relying on a polynomial 
approximation of some negative power of the Dirac operator.
The method presented in this section clearly shows that this problem,
even for a polynomial of very high degree $n$, is in practice
much less severe for the PHMC algorithm than for the multiboson
algorithm. This is a consequence of the fact that the number of
dynamical pseudofermion fields is $n$ in the multiboson algorithm
and only one in the PHMC.                                            
This allows in the latter case for a balance between the conflicting
requirements of minimizing the number of auxiliary pseudofermion fields
and maximizing the computational efficiency.

\subsection{Introducing more pseudofermion fields}

The last method of reducing memory requirements that we have studied
amounts to introducing more pseudofermion fields and distributing
the monomial factors of the polynomial $P_{n,\epsilon}(Q^2)$ among them.
Let us consider the action 
\be \label{S_P_multifields}
S_P^{(m)} = S_P^{(m)}[\phi,U] = 
\sum_{i=1}^m \phi_i^{\dagger} p_{n,\epsilon;m}^{(i)}(Q) \phi_i \; .
\ee
In eq.(\ref{S_P_multifields}) we have introduced $m$ 
positive definite subpolynomials $p_{n,\epsilon;m}^{(i)}(Q)$ each 
of degree $2n/m$ such that their product yields
$P_{n,\epsilon}(Q^2)$. In this way, one has 
to have memory space for only $ m + n/m $
pseudofermion fields in practise and hence would significantly reduce 
the memory requirements. 

However, it is clear already at this stage that by changing $m$ 
the dynamics of the algorithm will change: 
for $m=1$ we recover our PHMC algorithm. 
For $m=n$ we are in the case of
the original multiboson algorithm and would have an increase of the 
autocorrelation time with $n$. It might be hoped, however, that by choosing
$m$ small enough, the dynamics is not changed too much and that in this
way again a reduction of memory requirements can be achieved. 

It should be emphasized that when using the action
eq.(\ref{S_P_multifields}) special care has to be taken for the ordering
of the roots in order not to generate unwanted effects that come purely
from rounding errors. 
Without going into detail here, we note that by
using e.g. the bit-reversal scheme, a suitable ordering of the roots
avoiding rounding-error effects can be obtained. 
In addition, 
we checked that by running the
program with 64-bit precision our results, quoted below, did not change.

We have done several tests for
different choices of $m$ in the range $m\in [2,10]$.
We report our results obtained in the $SU(2)$ gauge
theory with two flavours of dynamical Wilson fermions. We set $c_{\rm sw}=0$
and take 
a lattice of size $8^3 \times 16$ with periodic boundary conditions.
We have considered two choices of the bare
parameters, $\beta=2.12$, $\kappa =0.15$ and $\beta=1.75$, $\kappa =0.165$,
using the subpolynomial and the bit-reversal ordering schemes of monomial factors. 

The effects for different choices of $m$ should most clearly appear in
the Hamiltonian 
\be \label{mol_dyn_ham}
H = \frac{1}{2} \sum_{x,\mu} \sum_{c=1}^{3} (\pi_{\mu}^c(x))^2
+ S_g[U] + S_P^{(m)}[\phi,U] \; ,
\ee
used in the molecular dynamics part of the PHMC algorithm. 
In eq.(\ref{mol_dyn_ham}) $\pi_{\mu}$ denote the momenta
conjugate to the gauge fields. 
We monitored the differences between the initial and final 
Hamiltonians in a molecular dynamics trajectory. 
For all parameters considered in table~\ref{tab:dyn_comp} we always started 
from the same thermalized gauge field configuration and kept the
step size $dt =0.04$ and the number of steps $N_{\rm step}=10$
fixed.

In table~\ref{tab:dyn_comp} we give our results
for the differences of the initial and final Hamiltonians 
$H_{\rm end}-H_{\rm in}$, of the gauge links $ \| U_{\rm end}-U_{\rm in} \| $ 
and of their conjugate momenta ($ \| \pi_{\rm end}-\pi_{\rm in} \| $), as measured
at the beginning and at the end of a trajectory. The definition 
of $ \| U_{\rm end}-U_{\rm in} \|^2 $ is analogous to the definition of
$\|U_{\rm in} - U_{\rm rev} \|^2 $, eq.(\ref{dstuff}), with a normalisation
factor of $(16V)^{-1}$ because the gauge group is now $SU(2)$. 
Finally, we define 
\be \label{mom_change}
\| \pi_{\rm end}-\pi_{\rm in} \|^2 = \frac{1}{12V}\sum_{x,\mu,c} | (\pi_{\mu}^c(x))_{\rm end}
- (\pi_{\mu}^c(x))_{\rm in} |^2 \; .
\ee

\begin{table*}[hbt]
\caption{The differences of the initial and final values of the Hamiltonian,
the momenta and the gauge links. Results are obtained 
on a $8^3 \times 16$ lattice at $\beta=1.75$, $\kappa =0.165$,
in the $SU(2)$ gauge theory.}                                              
\vspace{2mm}
\label{tab:dyn_comp}
\begin{tabular}{cccccc}
\hline
 $n$ & $\epsilon$ & $(m, {\rm order})$ & 
 $H_{\rm end}-H_{\rm in}$ & $ \| \pi_{\rm end}-\pi_{\rm in} \|^2 \quad \quad $ & 
 $ \| U_{\rm end}-U_{\rm in} \|^2 \quad \quad $  \\
\hline \hline
 $ 64   $ & $  0.0015$ & ($1$, BR) & 
 $ 0.63 $ & $0.301$ & $0.0657$  \\  
 $ 64   $ & $  0.0015$ & ($1$, SP) & 
 $ 0.63 $ & $0.301$ & $0.0657$  \\  
 $ 64   $ & $  0.0015$ & ($8$, BR) & 
 $ 28.1 $ & $1.170$ & $0.0357$  \\  
 $ 64   $ & $  0.0015$ & ($8$, SP) & 
 $ 40.9 $ & $1.158$ & $0.0354$  \\  
\hline \hline
 $ 64   $ & $  0.0005$ & ($1$, BR) & 
 $ 1.33 $ & $0.310$ & $0.0655$  \\  
 $ 64   $ & $  0.0005$ & ($8$, BR) & 
 $ 101  $ & $0.856$ & $0.0222$  \\  
\hline \hline
\end{tabular}
\end{table*}

As the results shown in table~\ref{tab:dyn_comp} indicate, 
the behaviour of the molecular dynamics part of the algorithm 
looks such that in the case with $m=8$ one typically
gets {\em larger} time discretisation effects. 
This is clearly 
seen by the values of 
$H_{\rm end}-H_{\rm in}$. At the same time, the difference in the
momenta $ \|\pi_{\rm end}-\pi_{\rm in} \|$ becomes larger, too, while the
difference in the gauge links $\| U_{\rm end}-U_{\rm in} \|$ becomes {\em smaller}  
than in the case $m=1$. 
This might be explained by the fact that the gauge links are always normalized to SU(2)
matrices and that they counteract the behaviour of the momenta to render
the difference $H_{\rm end}-H_{\rm in}$ small. 
The results depend also on the
distribution of the monomial factors among the subpolynomials $p_{n,\epsilon;m}^{(i)}$,
eq.(\ref{S_P_multifields}), as the 
comparison between the bit-reversal and the subpolynomial cases shows. 
When reducing the value 
of $\epsilon$ we again find even more different results, as shown
by the last two lines of table \ref{tab:dyn_comp}.  
Tests performed with gauge group SU(3) and Schr\"odinger functional
boundary conditions revealed a similar qualitative behaviour. 

We conclude that,
in order to get a reasonable acceptance rate 
in the cases with $m>1$, one is forced to reduce the value of
$dt$ substantially, resulting in a higher cost of a simulation. 
It seems to us that the case $m=1$, i.e. the PHMC algorithm
is most efficient. 
Of course, we cannot exclude that there are other possibilities 
of choosing subpolynomials that give a reduction of 
memory requirements and do not worsen the dynamical behaviour of the
algorithm. 
On the other hand, the solution to the problem of memory
requirement discussed in section
5.2 appears to be already satisfactory.

%% file: conclusion
\section{Conclusions}

We gave in this paper a detailed description of the PHMC algorithm,
which relies on a combination of the HMC algorithm and the
multiboson technique to simulate dynamical fermions \cite{taka,frezzi}.
We discussed the computational cost of the algorithm, checked that
rounding-error effects that can appear are under control
and showed possible ways to reduce memory requirements.

The effects of the correction factor that is introduced to render the
algorithm exact, was studied in detail. Special emphasis was put
on the fact that the PHMC algorithm samples the configuration space very
differently compared to the most commonly used HMC algorithm.
In particular, some evidence was given that the region of gauge
configuration space characterized by the presence of low lying modes
of $Q^2$ is explored much better when using the PHMC algorithm.

Of course, it is important to compare 
the performance of the PHMC algorithm with the one of                    
the HMC algorithm. The work presented here lays the ground for  
such an investigation of
the performance of the PHMC algorithm on which we will report in a separate
publication \cite{phmc_perf}. 
There we will also show further evidence that the PHMC algorithm 
samples configuration space differently from the HMC algorithm
and discuss consequences for physical observables.

\vspace{0.5cm}
{\large\bf Acknowledgements}

This work is part of the ALPHA collaboration research programme.
We are most grateful to S. Aoki, B. Bunk, R. Sommer, P.Weisz and U. Wolff
for many useful discussions and helpful comments.
In particular we thank U. Wolff for a critical reading of the manuscript
and M. L\"uscher for essential advices and discussions. 
We thank
DESY for allocating computer time to this project. R.F. thanks the
Alexander von Humboldt Foundation for the financial support for his
research stay at DESY--Hamburg, where part of this work was done.

\begin{appendix}

\section{Derivation of eq.(\ref{W_B_eval}) }

We start again from the $n_f=2$ lattice QCD partition
function, eq.(\ref{phmc2partition}):
\bea \label{phmc3partition}
{\cal Z} &
 = & \int {\cal D}U{\cal D}\phd{\cal D}\phi
{\cal D}\eta^\dagger{\cal D}\eta
  \; W_{B}[\eta,U] W_{IR}[U] \; e^{- (S_g + S_P + S_\eta ) } \nn \\
  S_P & = & S_P[U,\phi]=\phd P_{n,\epsilon}(Q^2[U]) \phi \nn \\
  S_\eta & = & \eta^\dagger\eta  \; .
\eea
The splitting of the original correction factor $W$ into two parts,
an ``infrared'' part,
\be
W_{IR}[U] =  \; \prod_{\lambda_j \le \tilde{\epsilon}}
[ 1 + R_{n,\epsilon}(\lambda_j) ] \; = \det ( L_{n,\epsilon,\tilde{\epsilon}}[U] ) \; ,
\ee
and a ``bulk'' part,
\be \label{W_B_with_L}
W_B[\eta,U]  = \; \exp \left\{\eta^\dagger [ 1 - L_{n,\epsilon,\tilde{\epsilon}}
\cdot (Q^2\cdot P_{n,\epsilon})^{-1} ] \eta \right\} \; ,
\ee
follows in a natural, unbiased way from the introduction of an operator which
acts on pseudofermion fields and depends
on $\tilde{\epsilon}$, $n$, $\epsilon$ and the gauge configuration $U$:
\be \label{L_oper_def}
L = L_{n,\epsilon,\tilde{\epsilon}}[U] =
1 + \sum_j | \lambda_j \rangle \langle \lambda_j |
R_{n,\epsilon}(\lambda_j) \theta (\tilde{\epsilon} - \lambda_j)  \; .
\ee

Since the index $j$ runs over all the eigenvalues of $Q^2$, the operator $L$,
which can be diagonalised simultaneously with $Q^2$, has eigenvalues given by
$1+R_{n,\epsilon}(\lambda_j)$ for the modes of $Q^2$ with
$\lambda_j \le \tilde{\epsilon}$ and by $1$ otherwise.
Due to the properties of the relative fit error function
$R_{n,\epsilon}(\lambda) = \lambda P_{n,\epsilon}(\lambda) -1$, 
the operator $L$ is Hermitean
and strictly positive, if all the $\lambda_j$'s are strictly positive: a zero mode
of $L$ appears only in one--to--one correspondence with a zero mode of $Q^2$.
In particular, the operator $L$ has exactly the same infrared behaviour
as the operator $Q^2 P_{n,\epsilon}(Q^2) = 1 + R_{n,\epsilon}(Q^2)$, if modes
with $\lambda_j \le \tilde{\epsilon}$ are present. However, because of the
$\theta$ functions appearing in its definition, $L$ is not a smooth functional
of the (lattice) gauge field, in contrast with the operator $Q^2 P_{n,\epsilon}(Q^2)$.

It is then straightforward to show that eq.(\ref{W_B_with_L}) can be rewritten
in the form of eq.(\ref{W_B_eval}) by introducing the pseudofermion field vector
$ | \eta_{\perp} \rangle $ as in eq.(\ref{eta_perp}).

%% file: appendix_b
\section{Optimizing memory requirements}
 
We present here a general and flexible method for some optimal trading of CPU time
with memory requirement in the implementation of the PHMC algorithm, which turns 
out to be very convenient in practical simulations.

Suppose that we wish to use the PHMC algorithm with the polynomial
$P_{n,\epsilon}(Q^2)$, eq.(\ref{pol_prod_q}), where only the degree $n$ is 
relevant for the present discussion, following the implementation
described in Section 2.2. Suppose also that the
lattice size and the memory capacity of our computer are such that,
in addition to gauge fields, their conjugate momenta and other
working arrays, only $N+1$ pseudofermion fields can be stored. One of these must
necessarily be the ``dynamical'' field $\phi $ extracted from the probability
distribution $\exp{[- \phd P_{n,\epsilon}(Q^2) \phi]}$: so we are left 
with the possibility of using at most $N$ auxiliary pseudofermion fields
during the evaluation of $\delta S_P$, eq.(\ref{PHMC_force}).
Since for $N \ge n$ it is possible to use the method described in Section 2.2, 
we consider here only the case $N < n$, which corresponds to the situation
of a relatively small storage capacity.

We have already observed in Section 2.2 that the variation $\delta S_P$,
eq.(\ref{PHMC_force}), of the pseudofermion action $S_P$, is a sum of
$n$ terms. Each term depends only on two auxiliary fields, $\chi_{j-1}$,
$\chi_{2n-j}$ (and their complex conjugates), where $j$ is the index
over which the sum runs. We remark that in evaluating $\delta S_P$
it is convenient to compute first the term in the sum with $j=n$, then
the one with $j=n-1$, etc., down to the last with $j=1$. Indeed, in this
way, the auxiliary fields $\chi_l$, with $l \ge n$ are required
in the following, natural order: first $\chi_n$, then $\chi_{n+1}$, etc.,
up to $\chi_{2n-1}$. Given this situation, the basic idea of our method
is to divide the available storage space for $N$ auxiliary pseudofermion fields
into three parts:
\begin{itemize}
\item {\bf (a)} A fixed storage part, where only $M$ of the auxiliary
 fields $\chi_1$, \dots, $\chi_n$ should be stored; let us denote 
 them by $\chi_{i_1}$, \dots, $\chi_{i_M}$.
\item {\bf (b)} A first working space part, where $K$ pseudofermion
 fields can be stored; this storage space should be {\em large enough} to construct
 $\chi_{n-1}$ starting from $\chi_{i_M}$, as well as $\chi_{i_m-1}$
 starting from $\chi_{i_{m-1}}$, for all $m=1,2,\dots ,M-1$.
\item {\bf (c)} A second working space part, where $2$ pseudofermion fields
 can be stored; this allows, for any given $l \ge n$, to construct and store 
 the field $\chi_{l+1}$,
 starting from the field $\chi_l$, while keeping it stored, too.
\end{itemize}

The relation $M+K+2 \le N$ must of course be satisfied\footnote{We leave
for the moment the freedom of using only part of the available memory, i.e.
taking $M+K+2 < N$.}. This relation and
the above requirement about the size of the part (b) of the storage space
impose restrictions on the possible choices of the integers $M$ and $K$,
as well as of the set of integers ${\cal I}_M \equiv \{i_1, i_2, \dots , i_M
\} $, satisfying $i_1 < i_2 < \dots < i_M < n$. For the moment, let us assume
that a choice of $M$, $K$ and ${\cal I}_M$ exists which satisfies our
requirements; we discuss below some examples and their practical performance.
It is clear that under these assumptions the evaluation of $\delta S_P$
can be performed following the strategy sketched in the steps below.

\begin{itemize}
\item  
In a preliminary step, starting from $\chi_0 \equiv \phi$, construct all
the auxiliary fields $\chi_j$, with $j \le i_M$ and store only $\chi_{i_1}$,
\dots, $\chi_{i_M}$ in the sector (a).

\item 
Set $i_{M+1}=n$, $i_0=0$. Then go through the following recursive procedure, where $s$ 
is an integer labelling the steps of the recursion: $s=1,2,\dots,M,M+1$.

\item 
For a given value of $s$, let us define the auxiliary integers $p=i_{M+1-s}$
and $q=i_{M+2-s}$. Then the step $s$ can be described as follows.
Starting from $\chi_{i_p}$ construct the fields
$\chi_{i_p+1}$, \dots , $\chi_{i_q-1}$
and store them in the sector (b). Then evaluate the contributions
to $\delta S_P$, eq.(\ref{PHMC_force}),
with $j=i_q,i_q-1, \dots , i_p+1$ (just in this order),
using the sector (c) to construct and temporarily store in turn the relevant
auxiliary fields $\chi_{2n-i_q} , \chi_{2n-i_q+1} , \dots , \chi_{2n-i_p-1}$.

\end{itemize}

It may happen that only part of the sector (b) has to be used in the
step with $s=M+1$. It is also important to remark that in the steps 
with $s>1$ a new evaluation of the auxiliary fields $\chi_j$ is required,
for all values of $j < i_M$ and not belonging to ${\cal I}_M$. Indeed, these
auxiliary fields were already constructed and then overwritten
during the preliminary step mentioned above. The CPU time needed for
their recomputation in the $M$ steps with $s>1$ represents 
the price to be paid for computing $\delta S_P$ using a number of
auxiliary fields less than the degree of the PHMC polynomial. On the
other hand, no such recomputation occurs in the step with $s=1$.

Let us come now to the determination of $M$, $K$, and ${\cal I}_M$ as functions
of $n$ and $N$. We recall that the chosen values of $M$ and $K$ must satisfy:
\be \label{memtrick_1a}
M+K+2 \le N \; , \quad \quad \quad \quad (M+1)(K+1) \ge n \; ,
\ee
where the second condition is equivalent to the above 
requirement on the size of the part (b) of the storage space. From
the description of our strategy for computing the variation of $S_P$, 
it should be clear that this condition guarantees that all of the 
$n$ terms appearing in eq.(\ref{PHMC_force}) for $\delta S_P$ 
can actually be evaluated. For any choice of $M$ and $K$ compatible
with eq.(\ref{memtrick_1a}), the set of integers ${\cal I}_M$ can
be defined as follows:
\be \label{memtrick_1b}
i_m = n - (M-m+1)(K+1) \; , \quad \quad \quad \quad m=1,2, \dots ,M \; .
\ee
On the other hand, with respect to the simple method of Section 2.2,
the computational overhead, due to the need of evaluating twice some
of the auxiliary fields, is given in units of $Q \phi$ operations by:
\be \label{memtrick_1c}
C_{\rm extra} = i_M - M = n-1-(K+M) \ge n-N+1 \; .
\ee
The optimal choices of $M$ and $K$ are the ones which minimize $C_{\rm extra}$,
{\em i.e.} maximize $M+K$, compatibly with eq.(\ref{memtrick_1a}): this
amounts to saturating the bound $M+K+2 \le N$ and yields $C_{\rm extra}=n-N+1$.
In table (\ref{tab:memtrick}) we illustrate the performance of this method 
for evaluating $\delta S_P$ in some typical cases. Moreover, that table
also contains results obtained by using a  
modified version of the method,
which is useful in cases when a very limited storage space is available.

This modified version relies on the fact that it is not strictly
necessary to keep constant, during the various steps of the computation
of $\delta S_P$, the size of the parts (a) and (b) of the storage space.
Indeed, we may only require that the parts (a) and (b) have a constant
{\em global} size, measured by the sum $M+K$. 
We will see that the freedom of varying the size of the single
parts (a) and (b) allows for reducing the minimal storage required for 
the computation of $\delta S_P$.

For instance, after the step with $s=1$ the auxiliary field $\chi_{i_M}$ is
no longer needed; which means that in the step with $s=2$ we may take the
parts (a) and (b) to have size $M-1$ and $K+1$, respectively. For the
same reason, after each step we may decrease by one unit the size of the
part (a) and increase by one unit the size of the part (b), ending
with a part (b) of size $K+M$ in the step with $s=M+1$. It is then
clear that, with a suitable definition of the integers in ${\cal I}_M$, 
the computation of $\delta S_P$ can be performed following exactly
all the steps of our method, as explained above. Of course, the meaning
of the integers $M$ and $K$ is now different: they only give the size
of the parts (a) and (b) during the preliminary step and the step with $s=1$.

The conditions to be fulfilled by the admissible choices of $M$ and $K$,
as functions of $n$ and $N$,
are also modified, as well as the corresponding definition of ${\cal I}_M$.
While the condition $M+K+2 \le N$ remains obviously valid, the definition
(\ref{memtrick_1b}) and the second condition in eq.(\ref{memtrick_1a})
should be replaced, respectively, by the recursive definition:
\begin{eqnarray} \label{memtrick_2b}
i_{M+1} & = & n  \nonumber  \\
i_{M+1-s} & = & i_{M+2-s} -s -K \; , \quad \quad \quad \quad
s = 1,2, \dots ,M+1 \\
\end{eqnarray}
and by the condition
\be \label{memtrick_2a}
i_1 \le K+M+1 \; .
\ee

On the other hand, the computational overhead with respect to the 
simple method of Section 2.2 can be shown to be still given by 
eq.(\ref{memtrick_1c}). As a consequence, the optimal choices of
$M$ and $K$ are the ones that maximize the sum $M+K$ and are compatible
with the new conditions necessary for the full evaluation of $\delta S_P$.
In the following we will refer to the two versions
of the method presented in this section, the one with
fixed size of the parts (a) and (b) of the memory and the 
one with variable size of the same parts, as to the basic and 
the modified version, respectively. The two versions are compared
in table (\ref{tab:memtrick}), for a few values of $n$ in the range
relevant for current day simulations using the PHMC algorithm. In
order to give an idea of the criticality of the simulations, we also
quote the typical condition number of $Q^2$, denoted by $k(Q^2)$, 
for the values of $n$ considered. 

For each value of $n$ , we consider different values of $N$: the minimal
one ($N=N_{min}'$) needed for using the modified version, the minimal one 
($N=N_{min}$) needed for using the basic version, $N=n/2$ (in order to
compare with the method of Section 5.1) and $N=n-N_{min}'$. In table 
(\ref{tab:memtrick}) a prime is used to denote the quantities relative
to the modified version of the computational method under study.

\begin{table*}[hbt]
\caption{Performance for the two versions of the method of Section 5.2
for computing $\delta S_P$, eq.(\ref{PHMC_force}). The notation
is defined in the text: $(M_0,K_0)$ denote, in the basic version of the method,
the choice of 
$(M,K)$ that minimizes, for given $n$ and $N$, both $C_{\rm extra}$
and $M$ itself. The corresponding 
primed quantities are relative to the modified version. Note that 
the minimal value of $C_{\rm extra}$ corresponding to the given
values of $n$ and $N$ is always realized.}
\vspace{2mm}
\label{tab:memtrick}
\begin{tabular}{lllllllll}
\hline
 $n$ & $k(Q^2)$ & $N $ & 
 $(M_0,K_0)$ & $C_{extra}$ & $C_{extra}/3n$ &
 $(M_0',K_0')$ & $C_{extra}'$ & $C_{extra}'/3n$ \\
\hline \hline
 $ 50   $ & $  770$ & $11$ & 
 $ \dots $ & \dots & \dots & 
 $ (7,2)$ & $40$ & $0.267$  \\  
 $ 50   $ & $  770$ & $15$ & 
 $ (4,9)$ & $36$ & $0.240$ & 
 $ (7,2)$ & $40$ & $0.240$  \\  
 $ 50   $ & $  770$ & $25$ & 
 $(2,21)$ & $26$ & $0.173$ & 
 $(2,21)$ & $26$ & $0.173$  \\  
 $ 50   $ & $  770$ & $39$ & 
 $(1,36)$ & $12$ & $0.080$ & 
 $(1,36)$ & $12$ & $0.080$  \\  
\hline \hline
 $  100  $ & $ 1470$ & $15$ &
 $ \dots $ & \dots & \dots &
 $ (11,2)$ & $86$ & $0.287$  \\
 $  100  $ & $ 1470$ & $20$ &
 $  (9,9)$ & $81$ & $0.270$ &
 $ (6,12)$ & $81$ & $0.270$  \\
 $  100  $ & $ 1470$ & $50$ &
 $ (2,46)$ & $51$ & $0.170$ &
 $ (2,46)$ & $51$ & $0.170$  \\
 $  100  $ & $ 1470$ & $85$ &
 $ (1,82)$ & $16$ & $0.053$ &
 $ (1,82)$ & $16$ & $0.053$  \\
\hline \hline
 $  180  $ & $ 4760$ &  $20$ &
 $ \dots $ & \dots & \dots &
 $ (14,4)$ & $161$ & $0.298$  \\
 $  180  $ & $ 4760$ &  $27$ &
 $(11,14)$ & $154$ & $0.285$ &
 $ (7,18)$ & $154$ & $0.285$  \\
 $  180  $ & $ 4760$ &  $90$ &
 $ (2,86)$ & $ 91$ & $0.169$ &
 $ (2,86)$ & $ 91$ & $0.169$  \\
 $  180  $ & $ 4760$ & $160$ &
 $(1,157)$ & $ 21$ & $0.039$ &
 $(1,157)$ & $ 21$ & $0.039$  \\
\hline \hline
\end{tabular}
\end{table*}

From table (\ref{tab:memtrick}) we can see that the method presented
in this section indeed enables one to save a significant amount of
storage space at the price of a very moderate computational overhead.
Namely, memory requirements are reduced by a large factor, which increases with
$n$ and is about $5 \div 9$ for the considered values of $n$.
On the other hand, the relative computational overhead, as measured by 
$C_{\rm extra}/3n$, eq.(\ref{memtrick_1c}), increases very slowly with $n$, 
approaching asymptotically the value $1/3$.

It is also important to remark that these numbers refer to the 
maximal memory saving that can be achieved by the method. However
we can see from table (\ref{tab:memtrick}) that for each value of $n$
many other choices of $N$ are allowed, which correspond to a different
balance between memory saving and computational efficiency. Such a flexibility
makes it very easy to optimize the balance between memory saving and
computational efficiency in any simulation setup,
which represents a clear advantage of the method presented in this section
in comparison with the ones discussed in Sections 2.2 and 5.1.

Let us come now to the comparison between the basic and the modified
version of the method presented in this section. 
The latter version turns out to be more
effective than the former in saving memory, as expected: we always
find $N_{\rm min}' \le N_{\rm min}$ in table (\ref{tab:memtrick}), 
where the dots stand for cases when the basic version does not allow
for a full evaluation of $\delta S_P$. However, for values of $N$
allowed in both versions, there is no difference in the computational
overhead between the two versions. Finally, we remark that for 
given values of $n$ and $N$, there are several choices of $M$ and $K$
that yield the minimal overhead ($C_{\rm extra} = n-N+1$) and are
compatible with all the necessary conditions specified above. At
this level, some further differences between the two versions appear, which
are however irrelevant in practice. In both versions, in particular,
it is not possible to choose a too small value of $M$, for given values
of $n$ and $N$, if all the necessary conditions are to be fulfilled
and the minimal overhead is to be achieved: the table (\ref{tab:memtrick})
shows also the lowest allowed values of $M$, and the corresponding ones
of $K$, for the different cases.

\end{appendix}